\documentclass[modern]{aastex61}

\usepackage{wasysym}

\submitjournal{ApJ} \shorttitle{ The $^{13}$C+$\alpha$ in AGB
stars} \shortauthors{Cristallo et al.}

\begin{document}

\title{The importance of the $^{13}$C($\alpha$,\MakeLowercase{n})$^{16}$O reaction in Asymptotic Giant Branch stars}

\correspondingauthor{Sergio Cristallo}
\email{sergio.cristallo@inaf.it}

\author{S. Cristallo}
\affiliation{INAF - Osservatorio Astronomico d'Abruzzo, via M.
Maggini snc, Teramo, Italy } \affiliation{INFN - Sezione di
Perugia, Via A. Pascoli snc, Perugia, Italy}

\author{M. La Cognata}
\affiliation{INFN - Laboratori Nazionali del Sud, Via S. Sofia 62, Catania, Italy}

\author{C. Massimi}
\affiliation{Dipartimento di Fisica e Astronomia, Universit\'a di Bologna, Via Irnerio 46, Bologna, Italy}
\affiliation{INFN - Sezione di Bologna, Viale Berti Pichat 6/2, Bologna, Italy}

\author{A. Best}
\affiliation{Universit\'a degli Studi di Napoli "Federico II", Via
Cintia, Napoli, Italy} \affiliation{INFN - Sezione di Napoli, Via
Cintia, Napoli, Italy}

\author{S. Palmerini}
\affiliation{Universit\'a degli Studi di Perugia,
Via A. Pascoli snc, Perugia, Italy}\affiliation{INFN - Sezione di Perugia, Via A. Pascoli snc,
Perugia, Italy} 

\author{O. Straniero}
\affiliation{INAF - Osservatorio Astronomico d'Abruzzo, via M.
Maggini snc, Teramo, Italy } \affiliation{INFN - Laboratori
Nazionali del Gran Sasso, Via G. Acitelli 22, Assergi, Italy}

\author{O. Trippella}
\affiliation{INFN - Sezione di Perugia, Via A. Pascoli snc,
Perugia, Italy}

\author{M. Busso}
\affiliation{Universit\'a degli Studi di Perugia, Via A. Pascoli
snc, Perugia, Italy} \affiliation{INFN - Sezione di Perugia, Via
A. Pascoli snc, Perugia, Italy}

%
\author{G.F. Ciani}
\affiliation{Gran Sasso Science Institute, Viale Francesco Crispi 7, L'Aquila, Italy }\affiliation{INFN - Laboratori Nazionali del Gran Sasso, Via G. Acitelli 22, Assergi, Italy}

\author{F.Mingrone}
\affiliation{CERN, Route de Meyrin 1211, Gen\'eve, France}

\author{L. Piersanti}
\affiliation{INAF - Osservatorio Astronomico d'Abruzzo, via M.
Maggini snc, Teramo, Italy } \affiliation{INFN - Sezione di
Perugia, Via A. Pascoli snc, Perugia, Italy}

\author{D. Vescovi}
\affiliation{Gran Sasso Science Institute, Viale Francesco Crispi
7, L'Aquila, Italy } \affiliation{INFN - Sezione di Perugia, Via
A. Pascoli snc, Perugia, Italy}

\begin{abstract}

Low mass Asymptotic Giant Branch stars are among the most
important polluters of the interstellar medium. In their
interiors, the main component (A$\gtrsim 90$) of the slow neutron
capture process (the $s$-process) is synthesized, the most
important neutron source being the $^{13}$C$(\alpha,$n$)^{16}$O
reaction. In this paper we review its current experimental status,
discussing possible future synergies between some experiments
currently focused on the determination of its rate. Moreover, in
order to determine the level of precision needed to fully
characterize this reaction, we present a theoretical sensitivity
study, carried out with the FUNS evolutionary stellar code and the
NEWTON post-process code. We modify the rate up to a factor of two
with respect to a reference case. We find that variations of the
$^{13}$C($\alpha$,n)$^{16}$O rate do not appreciably affect
$s$-process distributions for masses above 3 M$_\odot$ at any
metallicity. Apart from a few isotopes, in fact, the differences
are always below 5\%. The situation is completely different if
some $^{13}$C burns in a convective environment: this occurs in
FUNS models with M$<$3 M$_\odot$ at solar-like metallicities. In
this case, a change of the $^{13}$C($\alpha$,n)$^{16}$O reaction
rate leads to non-negligible variations of the elements surface
distribution (10\% on average), with larger peaks for some
elements (as rubidium) and for neutron-rich isotopes (as $^{86}$Kr
and $^{96}$Zr). Larger variations are found in low-mass
low-metallicity models, if protons are mixed and burnt at very
high temperatures. In this case, the surface abundances of the
heavier elements may vary by more than a factor 50.
\end{abstract}


\section{Introduction} \label{sec:intro}

As it is widely accepted, the main and the strong components of
the $s$-process (nuclei heavier than A$\sim 90$) found in the
solar system material have been produced by relatively low-mass
Asymptotic Giant Branch (AGB) stars ($1.2 <  $M/M$_\odot \le
4.0$), already extinct before the birth of the Sun. Their
structures consist of three layers: a degenerate C-O core, a thin
He-rich mantel (named He-intershell), and a loose and largely
convective H-rich envelope. They undergo recurrent He-shell
flashes (called thermal pulses, TPs) separated by relatively long
interpulse periods, during which a quiescent shell-H burning
balances the energy lost by radiation from the stellar surface. In
those objects, the most important neutron source is the
$^{13}$C$(\alpha,$n$)^{16}$O reaction, which is active in the
He-intershell. According to the current paradigm, a $^{13}$C
pocket forms at the beginning of each interpulse period in a small
layer characterized by a variable H-abundance profile. Then, as
this region contracts and warms up to $\approx 90-100$ MK,
$^{13}$C starts capturing $\alpha$ particles and, as a
consequence, releases neutrons. A second neutron burst, more rapid
than the first, but only marginally contributing to the
$s$-process nucleosynthesis, occurs during a TP, when the maximum
temperature in the convective zone powered by the He-flash exceeds
300 MK. In this case, neutrons are released through the
$^{22}$Ne$(\alpha,$n$)^{25}$Mg reaction. This neutron source
dominates the $s$-process nucleosynthesis of more massive AGB
stars ($4.0 <  M/M_\odot <  9.0$).

Starting from the early 80ies, different physical processes have been proposed as responsible for the synthesis of the $^{13}$C needed to reproduce observations. In particular, a thin transition zone containing a small amount of protons (only $10^{-6}$ M$_\odot$ of H) is needed in between the He-rich mantel and the H-rich envelope at its deepest penetration during a TDU episode. Then, at H re-ignition a $^{13}$C pocket rapidly forms. Such a $^{13}$C, which was originally thought to be engulfed in the convective shell triggered by the following TP (see, e.g., \citealt{ir82}), burns in radiative conditions during the long interpulse phase \citet{stra95}. As a matter of fact, the radiative $s$-process timescale is quite long (a few $10^4$ yr) and low neutron densities (namely $n_n<10^8$ cm$^{-3}$) are attained. Note that in spite of the low neutron density, the long timescale ensures a high enough neutron exposure to synthesize substantial $s$-process nuclei belonging to the main component. In the late 90ies, \citet{ga98} carried out a large number of nucleosynthesis post-process calculations. However, the mass and the profile of $^{13}$C within the pocket were treated as free parameters. In particular the {\it standard case}, corresponding to about $4\times10^{-6}$ M$_\odot$ of $^{13}$C, provided the best reproduction of the main $s$-process component in the solar system. Subsequent papers confirmed the above described scheme (e.g. \citealt{goriely,lugaro}).

The question of what physical mechanism leads to the formation of
the $^{13}$C pocket is still open. In the last 20 yr several
hypotheses  have been advanced. \citet{he97} (and also
\citealt{he00}), firstly proposed a convective overshoot operating
during the TDU. Basing on prescriptions derived from 2D
hydrodynamical calculations of stellar convection \citep{fr96},
they assumed that the convective mixing velocity decreases
exponentially below the convective border. In this way they
obtained $^{13}$C pockets with masses of the order of
$(2-4)\times10^{-7}$ M$_\odot$ (thus 10 to 20 times smaller than
the standard case defined by \citealt{ga98}). Later on,
\citet{la99} investigated the possibility of rotational induced
mixing, while \citet{dt03} analyzed the effect of a weak
turbulence induced by gravity waves (see also \citealt{ba16}). It
should be noted that all these models adopt a diffusion scheme to
treat the mixing of protons into the He-rich and C-rich zone,
independently on the engine of this mixing.\\
\citet{stra06} (see also \citealt{cri09a}), proposed a different
algorithm in which the degree of mixing scales linearly with the
mixing velocity, instead of quadratically as in the diffusion
scheme. Then, by adopting an exponential decrease of the
convective velocity, they showed that it is possible to produce
$^{13}$C pockets larger than that obtained in previous studies and
able to provide the required production of main component
$s$-process isotopes. Such a velocity profile drops as:
\begin{equation} \label{param}
v=v_{CE}\exp{\left(-\frac{\Delta r}{\beta H_P}\right)} \; ,
\end{equation}
where $\Delta$r is the distance from the convective boundary
defined by the Schwarzschild criterion, $v_{CE}$ is the velocity
at the formal convective boundary, $H_P$ is the pressure scale
height and $\beta$ is a free parameter (calibrated to $\beta=0.1$,
see \citealt{cri11}). We refer to this mixing scheme as
Exponentially-VElocity-Profile mixing (EVEP mixing). Later,
\citet{pi13} investigated the effects induced by rotation on the
evolution of the $^{13}$C pocket and the related $s$-process
nucleosynthesis. Note that the adoption of the EVEP scheme
facilitates the penetration of the convective envelope during TDU
episodes. As a consequence, the model experiences TDUs when the
mass of its H-exhausted core is lower (with respect to models
without the EVEP mixing). Then, models may become C-rich (C/O$>1$)
at lower surface luminosities. \cite{guanda} demonstrated that
such a result well fits with the observational luminosity function
of C-stars.
\\
More recently, \citet{nb14} advanced a new hypothesis about the
formation of the $^{13}$C pocket (see also \citealt{trippa,pa18}),
suggesting that dynamo-produced buoyancy of magnetize materials
could provide the necessary physical mechanisms to transport
protons from the H-rich envelope into the He-C rich mantel.
Magnetic instabilities also supply a sufficient transport rate
capable to explain the formation of a quite large zone with low
$^{13}$C concentration. In this scenario, the original poloidal
field of a rotating star generates a toroidal field of similar
strength developing various instabilities \citep{parker,spruit}
among which the buoyancy of magnetized structures \citep{schu}.
The dynamics of buoyant magnetized domains is strongly dependent
on the physics of the stellar environment and it is in general
very complex. However, \citet{nb14} have shown that below the
convective envelopes of AGB stars special conditions are held and
the fully MHD equations can be solved exactly. Then a simple
formula for radial component of the buoyancy velocity can be
obtained and it describes a fast transport mechanism (see the
original paper for details).

Besides the difficulties related to the adopted physical recipe,
models also deal with the uncertainties related to many input
quantities, among which nuclear reaction rates. The one affecting
the main neutron source, i.e. $^{13}$C$(\alpha,$n$)^{16}$O, plays
a relevant role. \citet{cri09a} (and \citealt{cri11}) found that
in some low-mass AGB models, $^{13}$C may not be fully consumed
during the interpulse. When this happens, the residual $^{13}$C is
engulfed into the convective zone powered by the incoming TP and
burns at higher temperature. This additional neutron burst affects
the compositions of isotopes in the neighborhood of critical
branchings. Those authors showed that this process provides an
excess of some radioactive nuclei, such as $^{60}$Fe, which has
been proved to be alive in the early solar system
\citep{td,moste}. Although such an occurrence is limited to the
early TP-AGB phase of low-mass high-metallicity models, a
variation of the $^{13}$C$(\alpha,$n$)^{16}$O reaction rate may
enhance or suppress such a process. The
$^{13}$C$(\alpha,$n$)^{16}$O reaction may also play an important
role during proton ingestion episodes (PIEs). These peculiar
events, possibly occurring in the early TP-AGB phase of low-mass
low-metallicity stars, are characterized by a protons engulfment
in the convective shell triggered by the first fully developed TP.
As a consequence of a PIE, an on-flight H-burning occurs, with
important consequences on the energetic stellar budget and on the
following (rich) $s$-process nucleosynthesis \citep{cri09b}.
During this event, the energy provided by the
$^{13}$C$(\alpha,$n$)^{16}$O reaction (plus the additional
contribution from the relative neutron capture) plays a key role.
Note that a similar nucleosynthesis may develop during a (very)
late He-shell flash or an AGB final thermal pulse (PG 1159
spectral class or Sakurai's objects; see e.g.
\citealt{wehe,herwig}).

In case of radiative $^{13}$C burning, the Gamow peak energy of the $^{13}$C$(\alpha,$n$)^{16}$O reaction for the relevant AGB
temperature is about 200 keV, which is well below the lower limit
so far reached by the direct measurements. In \S \ref{sec:expe} we report the state-of-the-art relative to its low-energy cross section that serves as input for a sensitivity study of the $^{13}$C$(\alpha,$n$)^{16}$O reaction and the related s-process nucleosynthesis in low-mass AGB stars. This is presented in \S
\ref{sec:theory}, where two different hypotheses about the formation of the $^{13}$C pocket, namely the EVEP mixing and the magnetic mixing have been taken under consideration. In \S \ref{future} we describe planned experiments which strive to improve our current knowledge. Finally, our conclusions are presented in \S \ref{sec:conclu}.

\section{State of the art} \label{sec:expe}

\begin{figure}[ht!]
\plotone{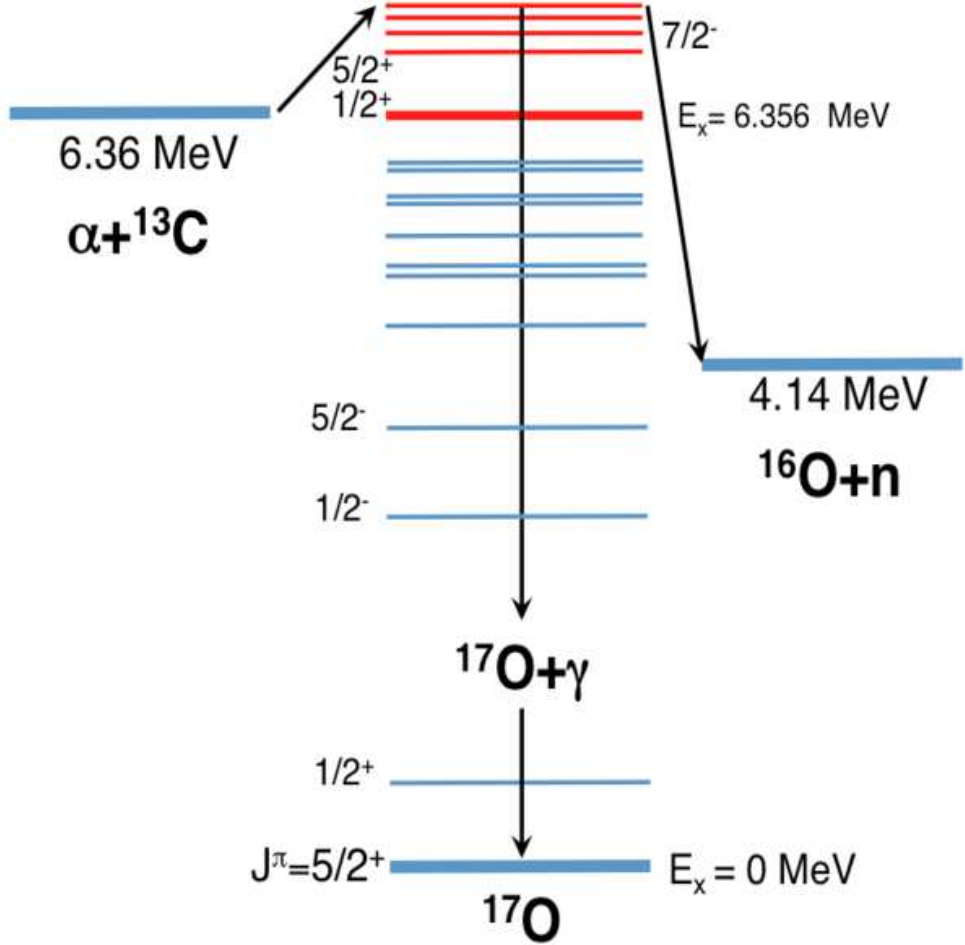} \caption{Schematic diagram
of the $^{13}{\rm C}(\alpha,$n$){}^{16}{\rm O}$ nuclear reaction
process, together with the competing exit channel
$^{17}$O+$\gamma$. Neutrons are produced via $\alpha$-particle
capture on $^{13}{\rm C}$ through a resonant process involving the
formation of the ${}^{17}{\rm O}$ compound nucleus. The excited
states of interest for AGB nucleosynthesis are shown in
red.}\label{fig:ReactionScheme}
\end{figure}

Owing to its astrophysical importance, the $^{13}{\rm
C}(\alpha,n){}^{16}{\rm O}$ S-factor\footnote{The astrophysical
S-factor is defined as: S(E)=E$\sigma$(E)exp(2$\pi\eta$), where
$\sigma$(E) is the cross section and $\eta$ is the Sommerfeld
parameter ($\eta$=Z$_1$Z$_2$e$^2$/$\hbar \nu$).} has been the
subject of many studies, aiming at the direct determination of its
cross section or focusing on specific $^{17}{\rm O}$ states.
A schematic diagram of the level scheme of $^{17}$O
is shown in Fig.~\ref{fig:ReactionScheme}. The $^{13}{\rm
C}+\alpha$ entrance channel and the two competing exit channels
$^{16}{\rm O}+n$ and $^{17}{\rm O}+\gamma$ are also represented by
black arrows. The ${}^{17}{\rm O}$ levels, near and above the
$\alpha-$threshold, of interest for AGB nucleosynthesis are marked
in red. 

Focusing on direct measurements, the most recent work by
\citet{HEI08} pointed out the substantial scatter of existing
data, showing a broad (up to a factor 2) range of absolute values
for the astrophysical S-factor.
On the other hand, the trend of the
astrophysical S-factor as a function of energy is consistent among
different data sets \citep{DAV68,BAI73,KEL89,DRO93,HAR05}. 
The lowest energy data point sits at a center-of-mass energy of
about 280~keV \citep{DRO93}, slightly above the Gamow window for
$\alpha$-induced $^{13}{\rm C}$ burning in radiative conditions.
Therefore,
extrapolation has proved necessary to supply a value of the
reaction rate at the temperatures of astrophysical interest. The
understanding of the low-energy behavior of the astrophysical
S-factor is complicated by the interplay between the rise in the
S-factor due to the excited state of $^{17}{\rm O}$ at
$E_x=6.356$~MeV \citep{ENDF} or $E_x=6.363$~MeV \citep{FAE15} with
spin parity $J^\pi =1/2^+$ (see Fig.~\ref{fig:ReactionScheme}) and
the enhancement produced by the electron screening effect
\citep{BRA90}. Moreover, such measurements are extremely
challenging since at $\sim 300$~keV the cross section is already
as low as $\sim 10^{-10}$~b and the neutron detection efficiency,
of about 30\%, further reduces the signal-to-noise ratio.

Therefore, indirect measurements turned out to be very useful to
constrain the $^{17}{\rm O}$ 6.356~MeV level contribution. 
These were essentially spectroscopic measurements of the
resonance energy \citep{FAE15}, of its squared Coulomb-modified
asymptotic normalization coefficient (ANC)
\citep{JOH06,AVI15}, and of the
corresponding spectroscopic factor
\citep{KUB03,KEE03,PEL08,GUO12,MEZ17}, which were used to
calculate the low-energy astrophysical S-factor and the $^{13}{\rm
C}(\alpha,$n$){}^{16}{\rm O}$ reaction rate. Concerning these last
measurements, aside from two conflicting cases
\citep{KUB03,JOH06}, very different experiments and analyses
supplied compatible values of ANCs, suggesting a minor
contribution of systematic errors, at odds with the
present status of direct measurements.

\subsection{Exploring the threshold region with THM} \label{asfin}

A different approach is used by the Trojan Horse Method (THM)
\citep{TRI14}.
Before 2015 \citep{FAE15}, it was believed that the near-threshold
$1/2^+$ $^{17}{\rm O}$ state was lying -3~keV below the $^{17}{\rm
O}\to{}^{13}{\rm C}+\alpha$ dissociation threshold \citep{TIL93}.
Therefore, the THM
turned out to be well suited to explore the energy interval where
such resonance dominates the astrophysical S-factor \citep{LAC10}. Indeed, this
approach allows one to bypass several drawbacks affecting
direct measurements, such as the steep drop characterizing the
cross section at energies far below the Coulomb barrier and the
electron screening enhancement of the astrophysical S-factor due
to atomic electrons in the target (about 20\% at $\sim300$~keV;
\citealt{DRO93}). 
Moreover, it offers the possibility to detect
charged particles instead of neutrons (possibly leading to
systematic uncertainties in the evaluation of the detection
efficiency).
In the THM framework, the $^{13}{\rm C}(\alpha,$n$){}^{16}{\rm O}$
S-factor was deduced by investigating the $^{13}{\rm
C}({}^{6}{\rm Li},$n${}^{16}{\rm O}){}^{2}{\rm H}$ process.
Then, $^6{\rm Li}$ binding energy and
$\alpha-d$ inter-cluster motion made it possible to reach
astrophysical energies in the $^{13}{\rm C}(\alpha,$n$){}^{16}{\rm
O}$ sub-reaction even if the THM reaction was induced at energies
of many MeV per nucleon. 
In early THM measurements \citep{LAC12,LAC13}, THM data were
scaled to the astrophysical S-factor recommended by \citet{HEI08} in
the $E_{^{13}{\rm C}-\alpha}$ region between $\sim0.6-1.2$~MeV.
As a results, a THM S-factor in good
agreement with the direct ones scaled to match the \citet{HEI08}
absolute value was attained in the $~0.28-1.2$~MeV energy region,
and a squared Coulomb-modified ANC for the $1/2^+$ $^{17}{\rm O}$
threshold state equal to $7.7 \pm
0.3_{stat}\,_{-1.5}^{+1.6}\,_{norm}\, {\rm fm}^{-1}$. This result
contradicts the existing independent assessments of the ANCs,
whose weighted average is $3.9\pm 0.5\, {\rm fm}^{-1}$ \citep{PEL08,GUO12,AVI15,MEZ17}.

\subsubsection{A concordance scenario for the $^{13}{\rm C}(\alpha,$n$){}^{16}{\rm O}$ S-factor}

Therefore, the pool of direct and
indirect data turned out to be incoherent. Such discrepancy could
not be reconciled taking into account the revised resonance
energy, setting its center at 4.7~keV above the $^{13}{\rm
C}-\alpha$ threshold \citep{FAE15}.
To reach a consistent S-factor, in a later THM work
\citep{TRI17} a change of paradigm was carried out, using the
existing ANC values to rescale the energy trend of the THM S-factor.
Discarding the normalization in
\citet{HEI08}, those authors concluded that only
\citet{BAI73,DRO93} supplied a direct data set compatible with the
THM S-factor and the ANC of the threshold level. With this new
normalization, a consistent ANC of $3.6\pm 0.7\, {\rm fm}^{-1}$
was obtained, in turn, from the THM data. Moreover, a THM S-factor
at E$_{^{13}{\rm C}-\alpha} = 140$~keV of $1.80_{-0.17}^{+0.50}
\times 10^6$~MeVb was deduced, to be compared with the
astrophysical factor by \citet{HEI08} ${\rm
S(140\,keV)}=2.2^{+1.1}_{-0.8}\times10^6$~MeVb.

These considerations show that, at present, the main drawback of
existing data is the absolute normalization, essentially connected
with neutron detection. Indirect measurements indicate that more
direct data are mandatory both at low energy, below about $\sim
0.3$~MeV, to deduce the electron screening potential, and at
higher energies, to supply a sound absolute normalization for
existing direct and indirect data. 

\section{Sensitivity study} \label{sec:theory}

Previous papers, devoted to the analysis of the
$^{13}$C($\alpha$,n)$^{16}$O reaction, mainly focused on the
effects induced by the adoption of rates proposed by different
authors. On the other hand, a sensitivity study is well suited to
evaluate the expected variations on the $s$-process, because it
allows the determination of relative distributions and, thus, the
corresponding derivatives with respect to variations of the rate
itself. Being the most recent direct measurement, we assume as
reference the rate proposed by \cite{HEI08} and we vary it
uniformly over the whole energy range by a factor 1.5. In such a
way we cover most of the rates presented in the literature so far.
The two most extreme cases, i.e. that proposed by \cite{KUB03} and
\cite{cf88}, are taken into account by varying the reference rate
by a factor of 2.

As already highlighted in \S \ref{sec:intro}, the identification
of the physical mechanism leading to the formation of the main
$^{13}$C reservoir in AGB stars (the so-called $^{13}$C pocket) is
a scientific issue debated since the end of the eighties. No
consensus has been reached on this topic to date, with different
processes proposed so far: opacity EVEP mixing
\begin{figure*}[htpb]
\centering
\includegraphics[width=\textwidth]{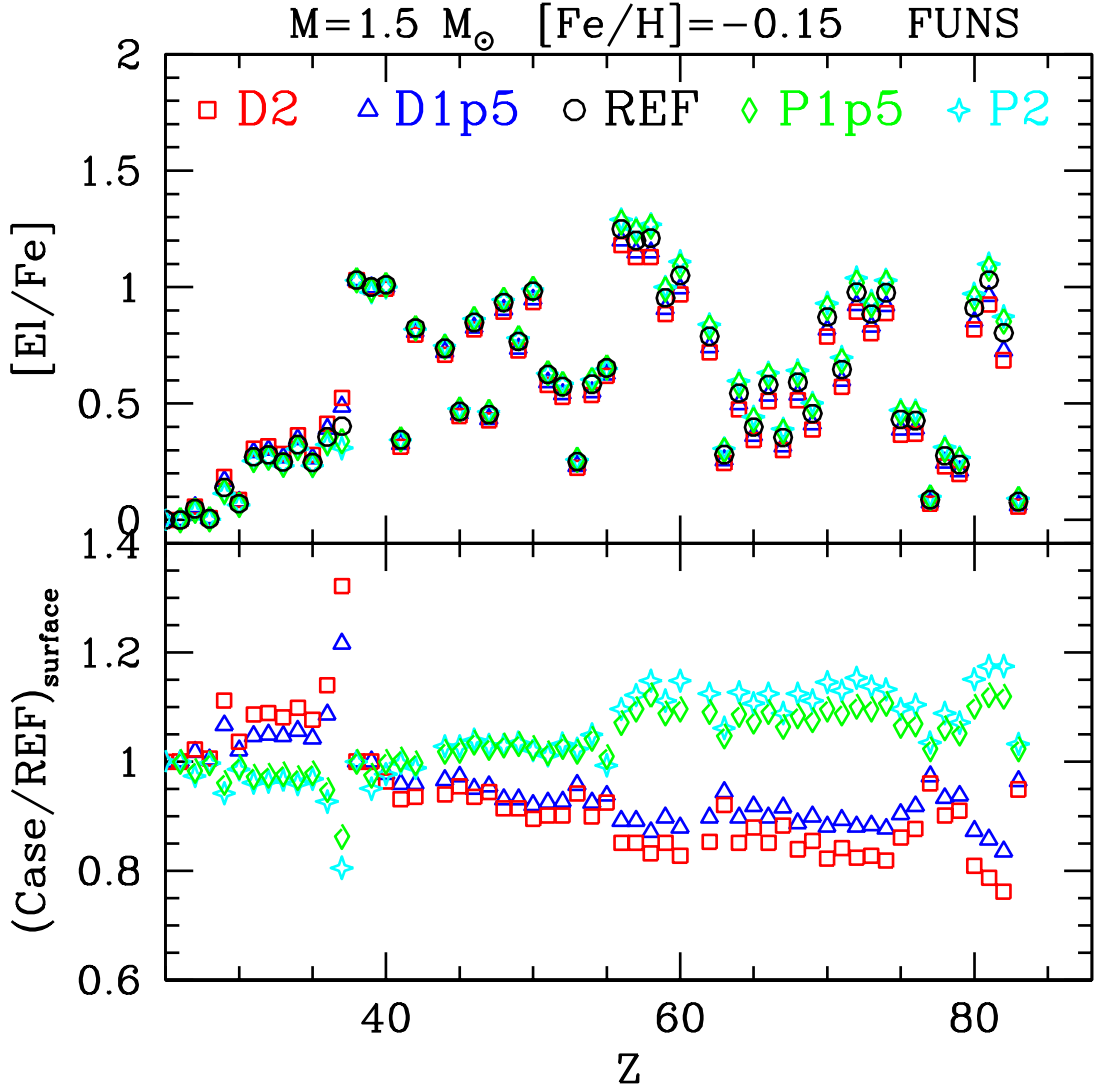}
\caption{FUNS heavy elements surface distribution
for an AGB star with initial mass M= 1.5 M$_\odot$ and
[Fe/H]=-0.15. Various symbols relate to different choices for the
$^{13}$C($\alpha$,n)$^{16}$O rate (see text for details).}
\label{fig_theo_1}
\end{figure*}
\begin{figure*}[htpb]
\centering
\includegraphics[width=\textwidth]{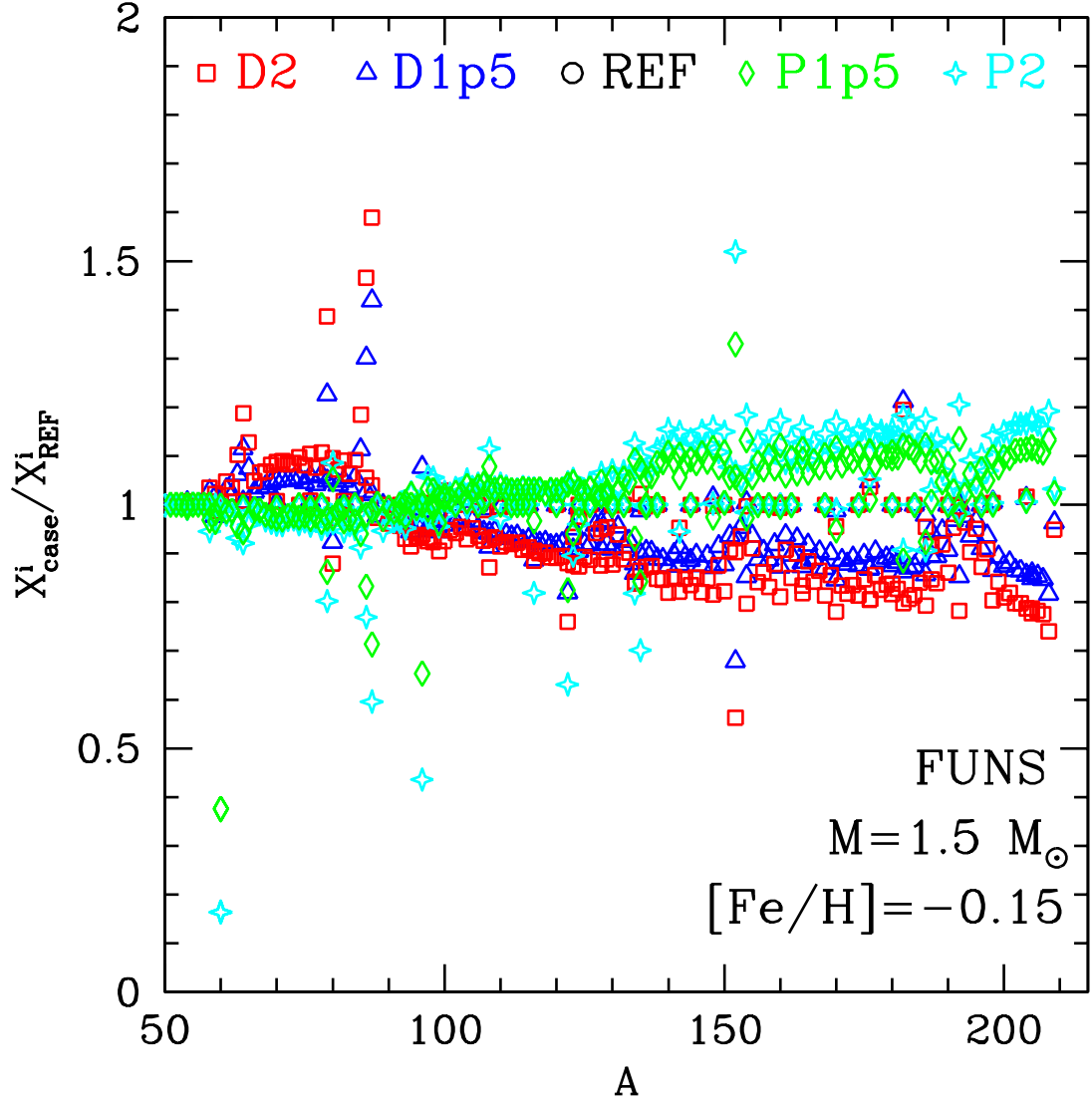}
\caption{As in Figure \ref{fig_theo_1}, but for the
isotopic composition.} \label{fig_theo_2}
\end{figure*}
\citep{stra06,cri09a}; magnetic buoyancy \citep{nb14,trippa,pa18};
Kelvin-Helmholtz instability coupled to gravity waves
\citep{ba16}. Those approaches lead to similar
(at first glance) $s$-process distributions; however, their comparison is not the main goal of this paper.\\
Hereafter, we concentrate on $s$-process nucleosynthesis
variations induced by the adoption of a modified
$^{13}$C($\alpha$,n)$^{16}$O rate in models with various masses
and metallicities, computed with different codes. We analyze two
sets of AGB models:
\begin{itemize}
\item{FUNS\footnote{\software{FUll Network Stellar \citep{stra06}}.}
evolutionary models: 1.5 M$_\odot$ and 3.0 M$_\odot$ with [Fe/H]=
-0.15 (corresponding to Z= $10^{-2}$, slightly lower than the
initial solar metallicity Z$_\odot=1.38\times 10^{-2}$), 4.0
M$_\odot$ with [Fe/H]= -2.15 (Z= $2.45\times 10^{-4}$, considering
an initial enrichment of $\alpha$-elements [$\alpha$/Fe]=0.5) and
1.3 M$_\odot$ with [Fe/H]= -2.85 (Z= $4.9\times 10^{-5}$,
considering an initial enrichment of $\alpha$-elements
[$\alpha$/Fe]=0.5);} \item{NEWTON post-process calculations on a
2.0 M$_\odot$ with [Fe/H]= -0.15 using AGB stellar structures
computed with the FRANEC\footnote{\software{Frascati RAphson-Newton
Evolutionary Code \citep{cs89}}.} code (i.e. adopting a pure
Schwarzschild criterion for the identification of convective
borders). Models of this kind were illustrated, e.g., by
\citet{stra03}.}\end{itemize}
\begin{figure*}[htpb]
\centering
\includegraphics[width=\textwidth]{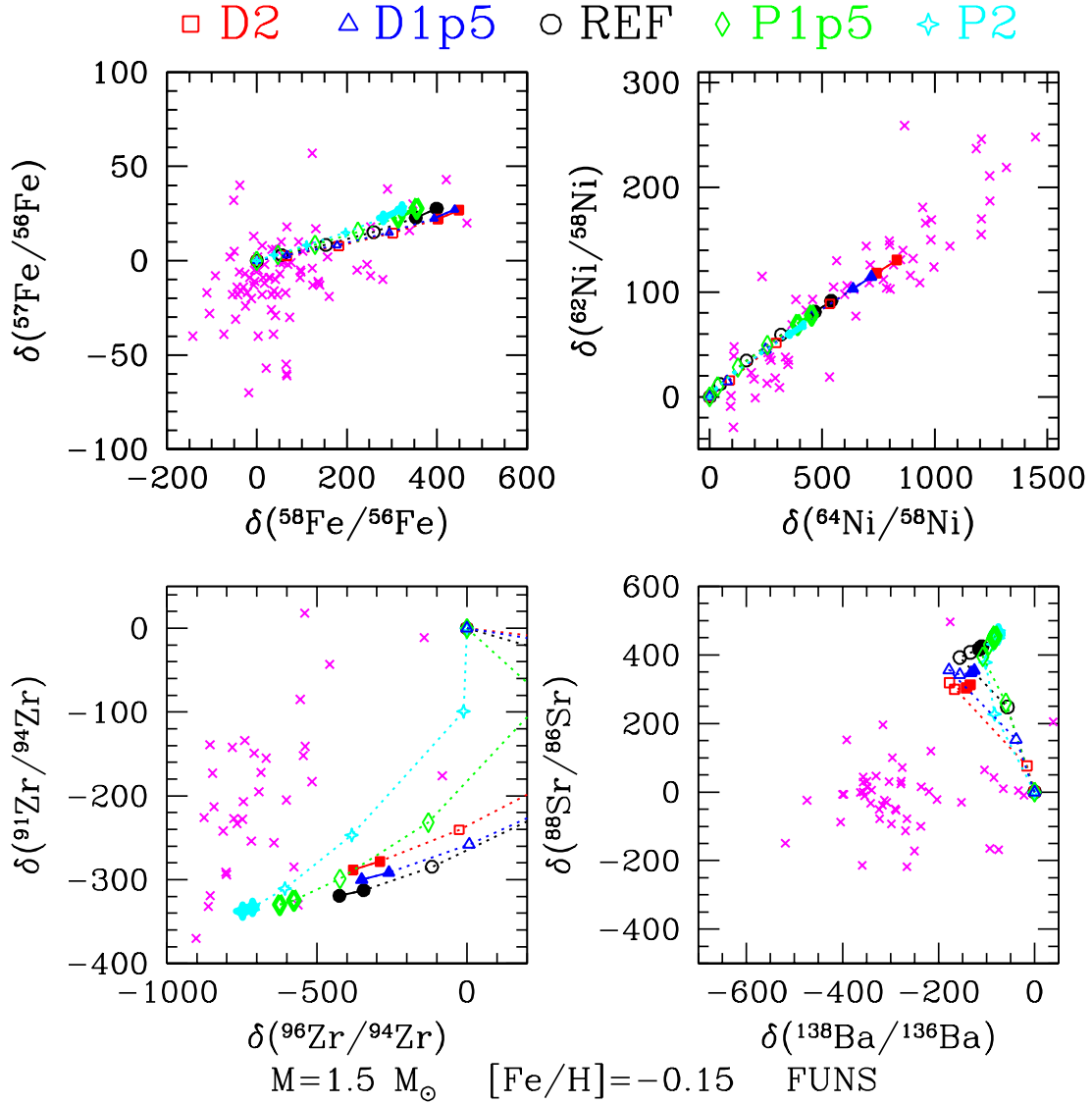}
\caption{Comparison between measured isotopic
anomalies for various elements, compared to a model with M= 1.5 M
$_\odot$, [Fe/H]=-0.15 and different choices for
the$^{13}$C($\alpha$,n)$^{16}$O rate (see text for details).}
\label{fig_theo_3}
\end{figure*}

\subsection{FUNS models}

A detailed description of FUNS models can be found in
\citet{cri16} and references therein. FUNS is derived from the
FRANEC code \cite{cs89}. Major improvements with respect to
previous versions of the code are the mass-loss law, the use of a
full nuclear network (from hydrogen to bismuth) directly coupled
to the physical evolution of the model, and, finally, the use of
an exponentially decaying profile of convective velocities at the
inner border of the envelope (see \citealt{stra06} for details).
As already reported in \S \ref{sec:intro}, the introduction of
such an algorithm has deep consequences on the physical and
chemical evolution of the model: it makes the border stable
against perturbations, it increases the efficiency of TDU
episodes, and it allows the formation of a tiny $^{13}$C-rich
region (the $^{13}$C pocket) at the base of the envelope after
each TDU. Note that the external region of the $^{13}$C pocket is
$^{14}$N-rich, due to the larger number of available protons at
the H-shell re-ignition. Such an isotope acts as a major neutron
poison via the $^{14}$N(n,p)$^{14}$C reaction, thus reducing the
number of neutrons available for the synthesis of heavy elements.
\begin{figure*}[htpb]
\centering
\includegraphics[width=\textwidth]{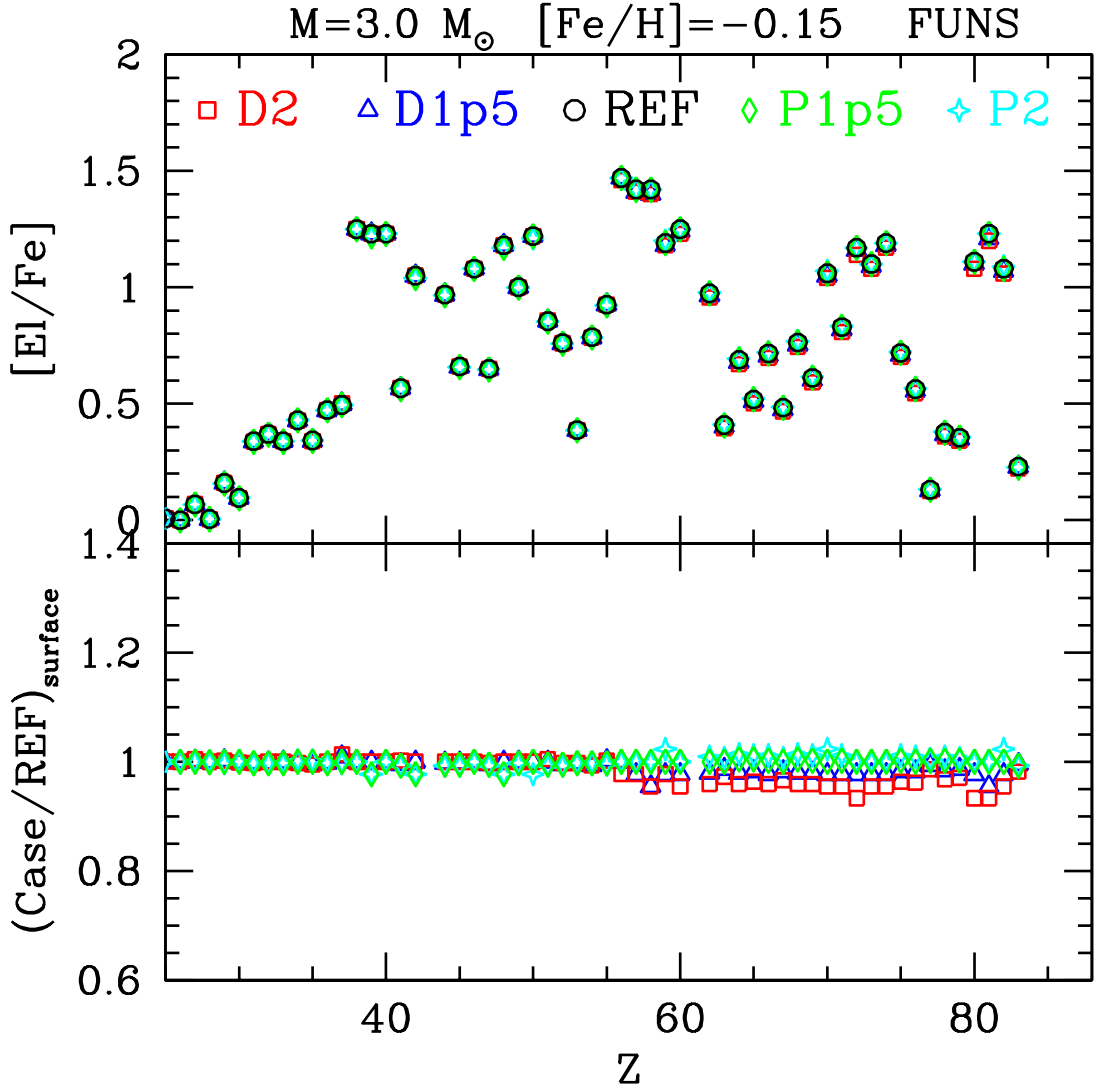}
\caption{As in Figure \ref{fig_theo_1}, but for a
star with initial mass M= 3 M$_\odot$.} \label{fig_theo_4}
\end{figure*}
\begin{figure*}[htpb]
\centering
\includegraphics[width=\textwidth]{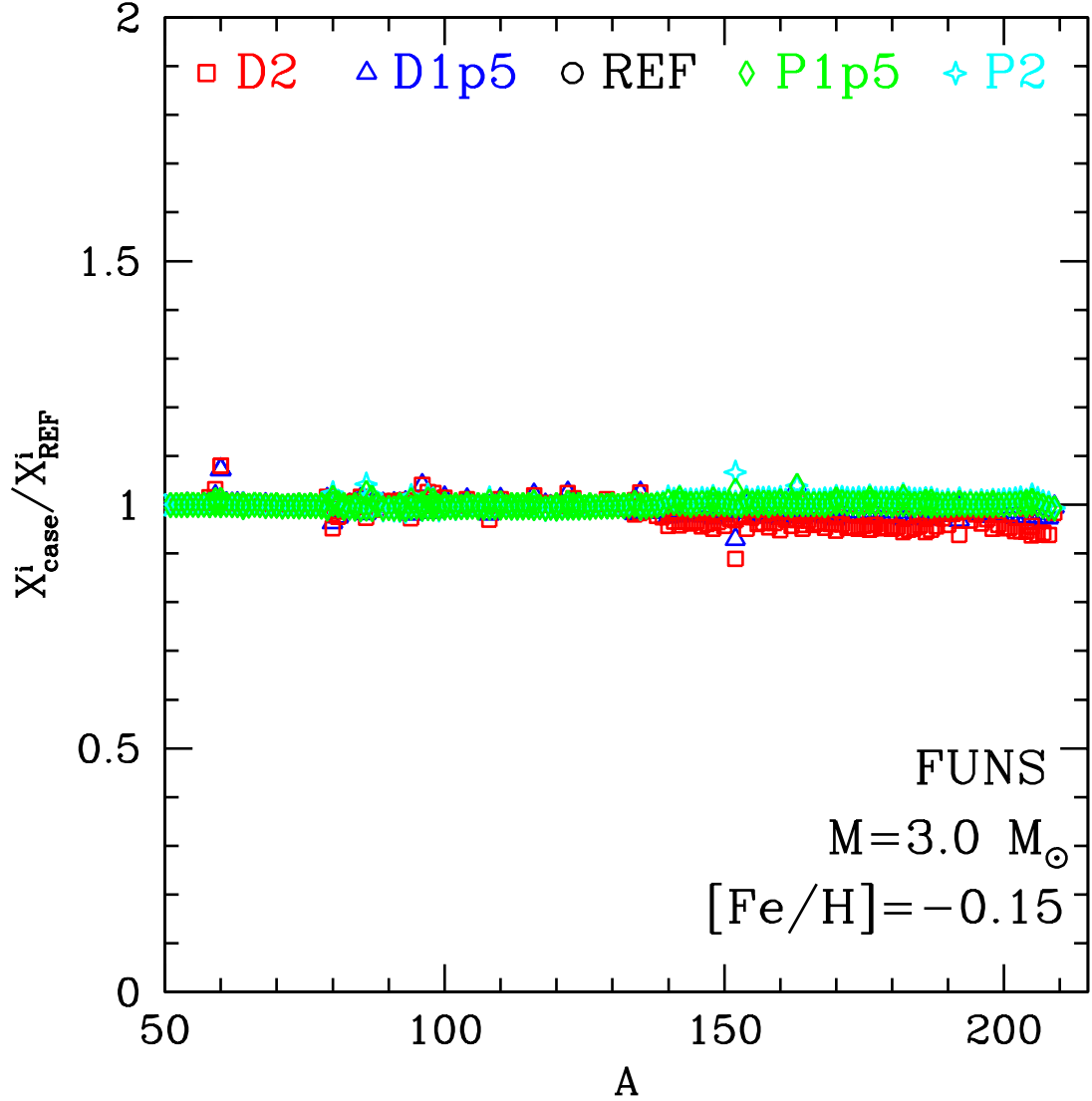}
\caption{As in Figure \ref{fig_theo_4}, but for the
isotopic composition.} \label{fig_theo_5}
\end{figure*}
\subsubsection{Solar-like metallicity}

In this section we discuss the effects induced by a change of the
$^{13}$C($\alpha$,n)$^{16}$O reaction rate in models of low mass
(1.5 M$_\odot$ and 3.0 M$_\odot$) and [Fe/H]= -0.15. This
metallicity is representative of the environment where pre-solar
SiC grains formed. Those $\micron$-sized particles are relics of
the nucleosynthesis in the interiors of already extinct AGB stars,
whose material polluted the Solar System before its formation
(see, e.g, \citealt{bgw}). The isotopic anomalies detected in
those grains are, in fact, clearly connected to physical
conditions not available in the Solar System at the epoch of its
formation.
\begin{figure*}[htpb]
\centering
\includegraphics[width=\textwidth]{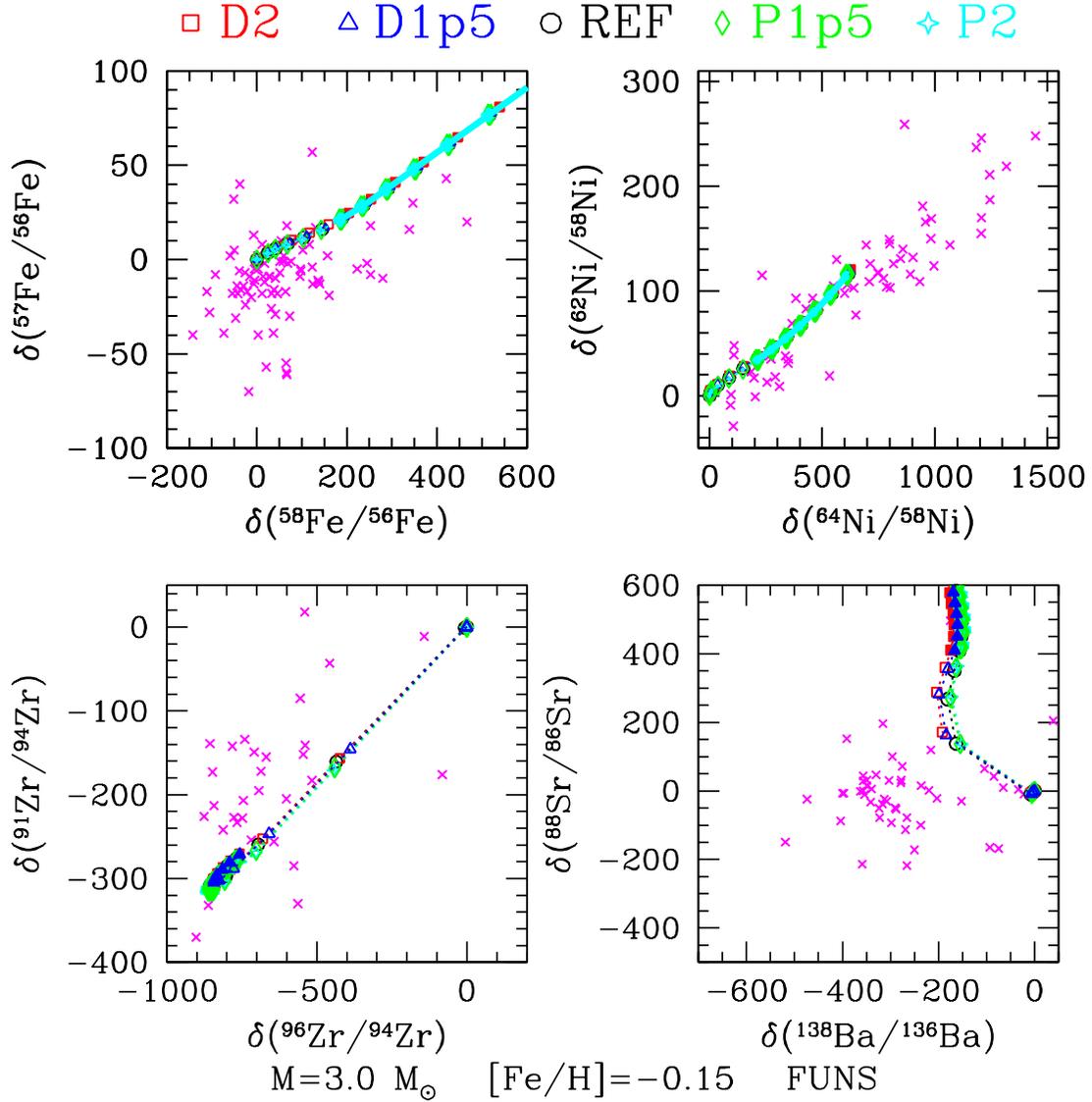}
\caption{As in Figure \ref{fig_theo_3}, but for a
star with initial mass M= 3 M$_\odot$.} \label{fig_theo_6}
\end{figure*}
In the upper panel of Figure \ref{fig_theo_1} we report the heavy
elements ($Z>25$) surface composition of a star with M= 1.5
M$_\odot$ and [Fe/H]=-0.15\footnote{The spectroscopic notation is
adopted:
[El/Fe]=log(N(El)/N(Fe))$_{star}$-log(N(El)/N(Fe))$_\odot$}.
Symbols refer to the different adopted rates for the
$^{13}$C($\alpha$,n)$^{16}$O reaction: reference case (hereafter
REF, \citealt{HEI08}; dark dots); reference case divided by a
factor 2 (hereafter D2; red squares); reference case divided by a
factor 1.5 (hereafter D1p5; blue triangles); reference case
multiplied by a factor 1.5 (hereafter P1p5; green diamonds);
reference case multiplied by a factor 2 (hereafter P2; cyan
stars). The corresponding isotopic surface distributions are
plotted in Figure \ref{fig_theo_2}. An inspection to the surface
elemental composition of the REF model reveals an almost flat
overproduction of the three $s$-process peaks (upper panel of
Figure \ref{fig_theo_1}), namely the ls component (Sr-Y-Zr), the
hs component (Ba-La-Ce-Pr-Nd) and lead. An increase of the
$^{13}$C($\alpha$,n)$^{16}$O reaction leads to a larger production
of hs elements ($\sim$15\%) and to an even slightly higher
synthesis of lead ($\sim$20\%)\footnote{Bismuth does not show a
comparable increase due to the very low neutron capture cross
section of $^{208}$Pb, which strongly limits $^{209}$Bi production
at solar-like metallicities}. On the contrary, elements lighter
than strontium are slightly under-produced. At first glance, this
result could appear in contrast to the fact that the $^{13}$C in
the pocket fully burns in radiative conditions between two TPs
\citep{stra95}. Actually, this is true for most of the AGB models,
but it does not hold for low mass stars (M$<$3.0 M$_\odot$) at
solar-like metallicities ([Fe/H]$\geq -0.15$).
\begin{figure*}[htpb]
\centering
\includegraphics[width=\textwidth]{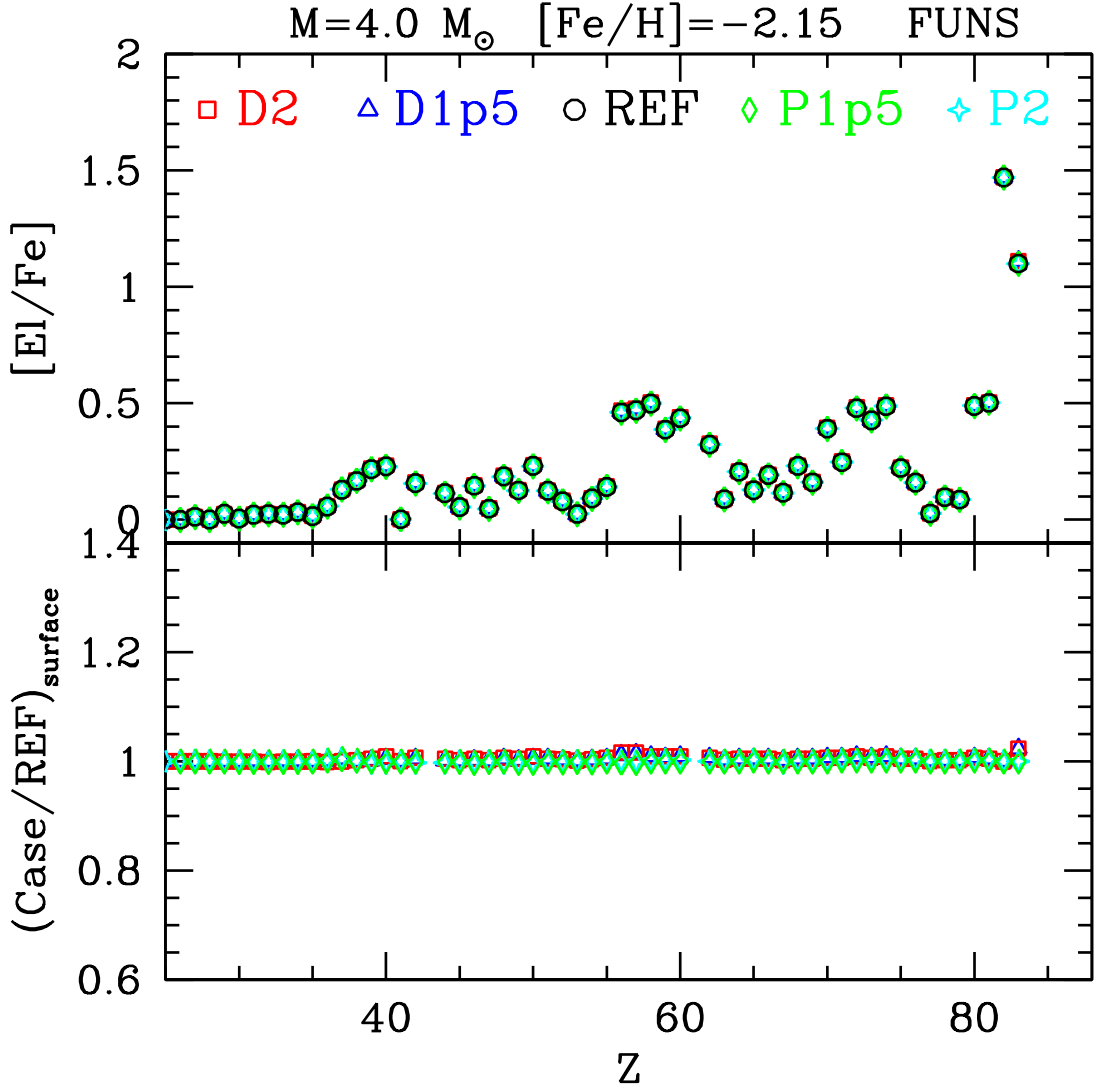}
\caption{As in Figure \ref{fig_theo_1}, but for a
star with initial mass M= 4 M$_\odot$ and [Fe/H]=-2.15.}
\label{fig_theo_7}
\end{figure*}
In fact, \citet{cri09a} demonstrated that, for those models, part
of the $^{13}$C in the first pockets is engulfed in the convective
shell generated by the following TP (see also \citealt{ka10}).
This derives from the fact that the $^{13}$C in the pocket does
not have enough time to fully burn in a radiative environment. AGB
models with larger initial masses, instead, have high enough
internal temperatures to guarantee a complete radiative $^{13}$C
burning before the onset of the following TP. Such a behavior is
easily understood by inspecting the masses of the H-exhausted
cores (M$_{\rm H}$) at the beginning of the TP-AGB phase. At
solar-like metallicities, stars with M$<$3 M$_\odot$ have almost
the same M$_{\rm H}$, while in more massive stars M$_{\rm H}$
linearly grows with the stellar mass (see, e.g. Figure 2 of
\citealt{cri15}). The larger M$_{\rm H}$ is, the larger the
temperature in the He-intershell region is. In addition, the lower
the metallicity is, the larger M$_{\rm H}$ is. Therefore, the
convective $^{13}$C burning disappears with increasing the stellar
mass and/or with decreasing the initial metal content. When some
$^{13}$C is ingested in the TP, it burns at a definitely higher
temperature, producing rather large neutron densities ($\sim
10^{11}$ cm$^{-3}$). During that episode, the production of
heavier elements is disfavored by convection, which does not allow
isotopes to locally pile up. Moreover, the abundant $^{14}$N in
the upper region of the $^{13}$C pocket acts as a neutron poison
(via the $^{14}$N(n,p)$^{14}$C reaction), thus further decreasing
the number of neutrons available for the nucleosynthesis of
elements heavier than iron. Another interesting feature of
convective $^{13}$C burning is that the production of neutron rich
isotopes close to $s$-process branchings largely increases with
respect to standard radiative $^{13}$C burning.
\begin{figure*}[htpb]
\centering
\includegraphics[width=\textwidth]{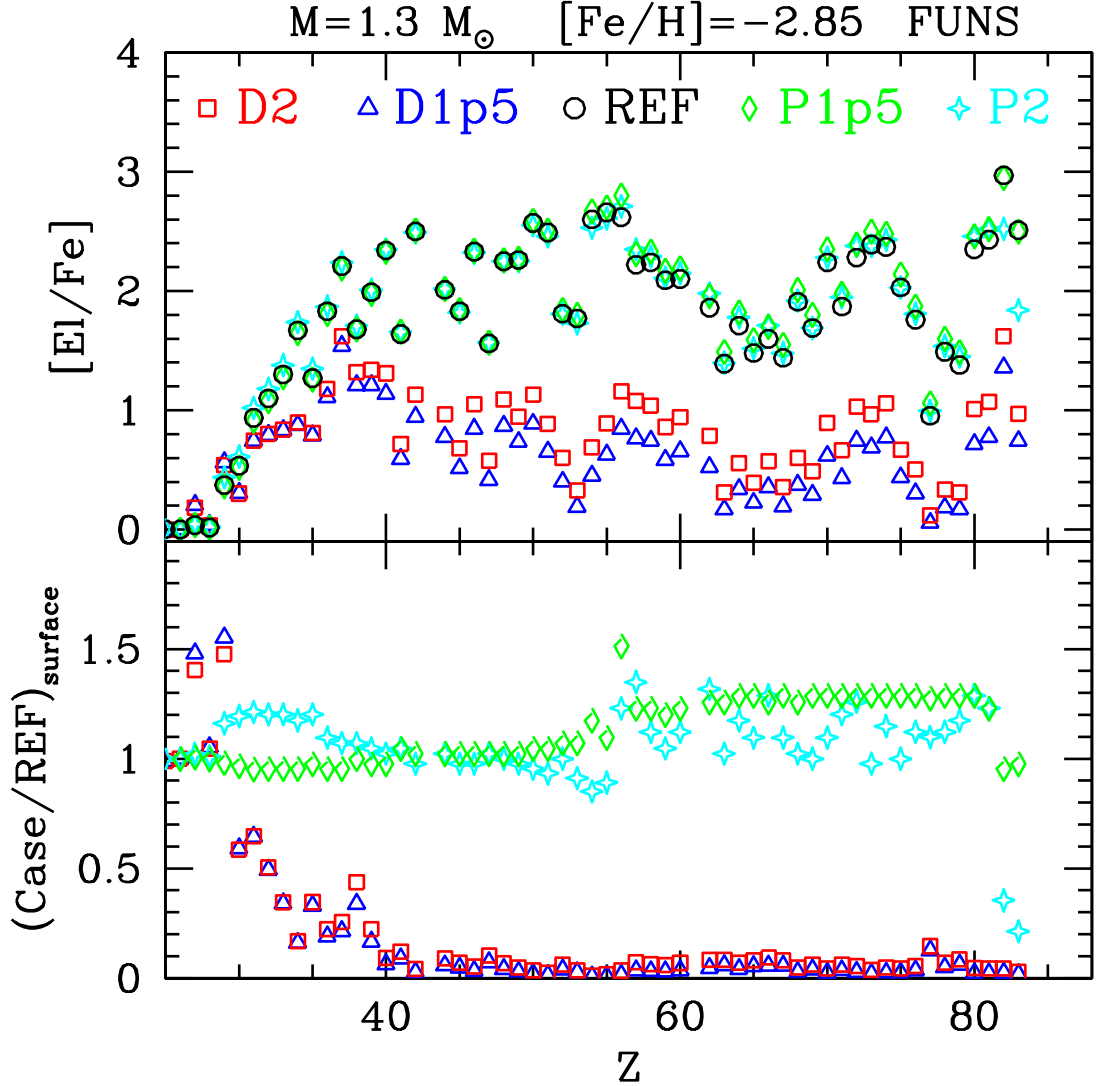}
\caption{As in Figure \ref{fig_theo_1}, but for a
star with initial mass M= 1.3 M$_\odot$ and [Fe/H]=-2.85.}
\label{fig_theo_8}
\end{figure*}
This is the case, for instance, for $^{60}$Fe (the largest
variation, almost a factor 20, is found for this isotope),
$^{86}$Kr, $^{87}$Rb and $^{96}$Zr (see Figure \ref{fig_theo_2}).
On the contrary, other isotopes are overproduced with a large
$^{13}$C($\alpha$,n)$^{16}$O rate. For example, the production of
$^{152}$Gd increases by more than 50\% when the rate is multiplied
by a factor 2. This is due to the fact that the nucleosynthesis of
such an isotope strongly depends on the $^{151}$Sm branching,
which is open during standard radiative $^{13}$C burning (thus
$^{152}$Gd is partly fed by the main $s$-process flow). On the
contrary, this branching is closed at high temperatures, as it
occurs during convective $^{13}$C burning\footnote{For an
interested reader, we refer to \citet{bi15}, where all branchings
of the $s$-process are described in detail.}. The increased
production of $^{87}$Rb characterizing the D1p5 and D2 cases leads
to the largest element surface variation, which is found for
rubidium (+30\%). In principle, the observed Rb/Sr ratio in C-rich
stars belonging to the Galactic disk could be used to constrain
the efficiency of the $^{13}$C($\alpha$,n)$^{16}$O reaction.
Unfortunately, the mass of those stars is poorly determined;
moreover, current observational uncertainties are of the same
order of magnitude of the theoretical differences just described.
Thus, from this analysis only extremely low values for the
$^{13}$C($\alpha$,n)$^{16}$O reaction can be safely discarded.
More precise hints on the efficiency of the
$^{13}$C($\alpha$,n)$^{16}$O reaction can be derived from the
analysis of isotopic ratios in pre-solar SiC grains. In Figure
\ref{fig_theo_3} we compare FRUITY models with available
laboratory measurements of selected key elements\footnote{Usual
meteoritic notation is used:
$\delta(^nX_Z)=((^nX_Z/^mX_Z)_{grain}/(^nX_Z/^mX_Z)_{Sun}-1)*1000$
\permil.}\citep{liu14a,liu14b,liu15,reto}. Note that we are
evaluating the effects induced by a change of the
$^{13}$C($\alpha$,n)$^{16}$O rate and not the capability of the
model to reproduce SiC data. The largest variations are found for
$^{64}$Ni and $^{96}$Zr. This is somewhat expected, being both
neutron-rich isotopes whose production depends
on an $s$-process branching (at $^{63}$Ni and $^{95}$Zr, respectively). \\
Finally, we highlight that the D2 and D1p5 models show
particularly large $^{86}$Kr/$^{82}$Kr ratios ($\sim$1) with
respect to the REF case (0.8), possibly giving hints on the
implantation energy of the $^{86}$Kr-rich component (see
\citealt{raut13}). In Table \ref{krzr} we report the
$^{86}$Kr/$^{82}$Kr and the $^{96}$Zr/$^{94}$Zr number ratios of
the computed models, together with their percentage variations
with respect to the reference case.
\begin{table}[ht!]
\renewcommand{\thetable}{\arabic{table}}
\centering \caption{$^{86}$Kr/$^{82}$Kr and $^{96}$Zr/$^{94}$Zr
number ratios of the 1.5 M$_\odot$ model with [Fe/H]=-0.15 for
different $^{13}$C($\alpha$,n)$^{16}$O rates, together with their
percentage variations with respect to the reference case.}
\label{krzr}
\begin{tabular}{lcc}
\tablewidth{0pt}
\hline
\hline

Case & $^{86}$Kr/$^{82}$Kr & $^{96}$Zr/$^{94}$Zr \\
\hline
D2 & 1.046 (+34\%)& 0.999 (+8\%)\\
D1p5 & 0.966 (+24\%)& 0.107 (+13\%)\\
REF & 0.777& 0.093\\
P1p5 & 0.669 (-14\%)& 0.061 (-35\%)\\
P2 & 0.633 (-19\%)& 0.041 (-56\%)\\
\hline
\end{tabular}
\end{table}

In Figures \ref{fig_theo_4}, \ref{fig_theo_5} and \ref{fig_theo_6}
we performed the same analysis already described, but for a star
with initial mass M=3 M$_\odot$ and the same metallicity ([Fe/H]=
-0.15). The heavy element distributions are almost
indistinguishable, apart from the D2 case, which is characterized
by a small (5\%) underproduction of the heaviest elements
(Z$>$50). The same holds for the isotopic composition, where the
only noticeable differences are found for $^{60}$Fe and $^{152}$Gd
(in any case below 10\%). The theoretical isotopic $\delta$ curves
do not evidence any appreciable deviation from the reference case.
This behavior confirms that, in more massive AGBs, the temperature
of the He-intershell region is always large enough to allow a
complete $^{13}$C burning during the radiative interpulse phase.
As highlighted before, this is a consequence of the larger core
mass of the 3.0 M$_\odot$ model at the first TP followed by TDU
(M$_{\rm H}\sim 0.59$ M$_\odot$), with respect to the 1.5
M$_\odot$ model (M$_{\rm H}\sim 0.56$ M$_\odot$).
\begin{figure*}[htpb]
\centering
\includegraphics[width=\textwidth]{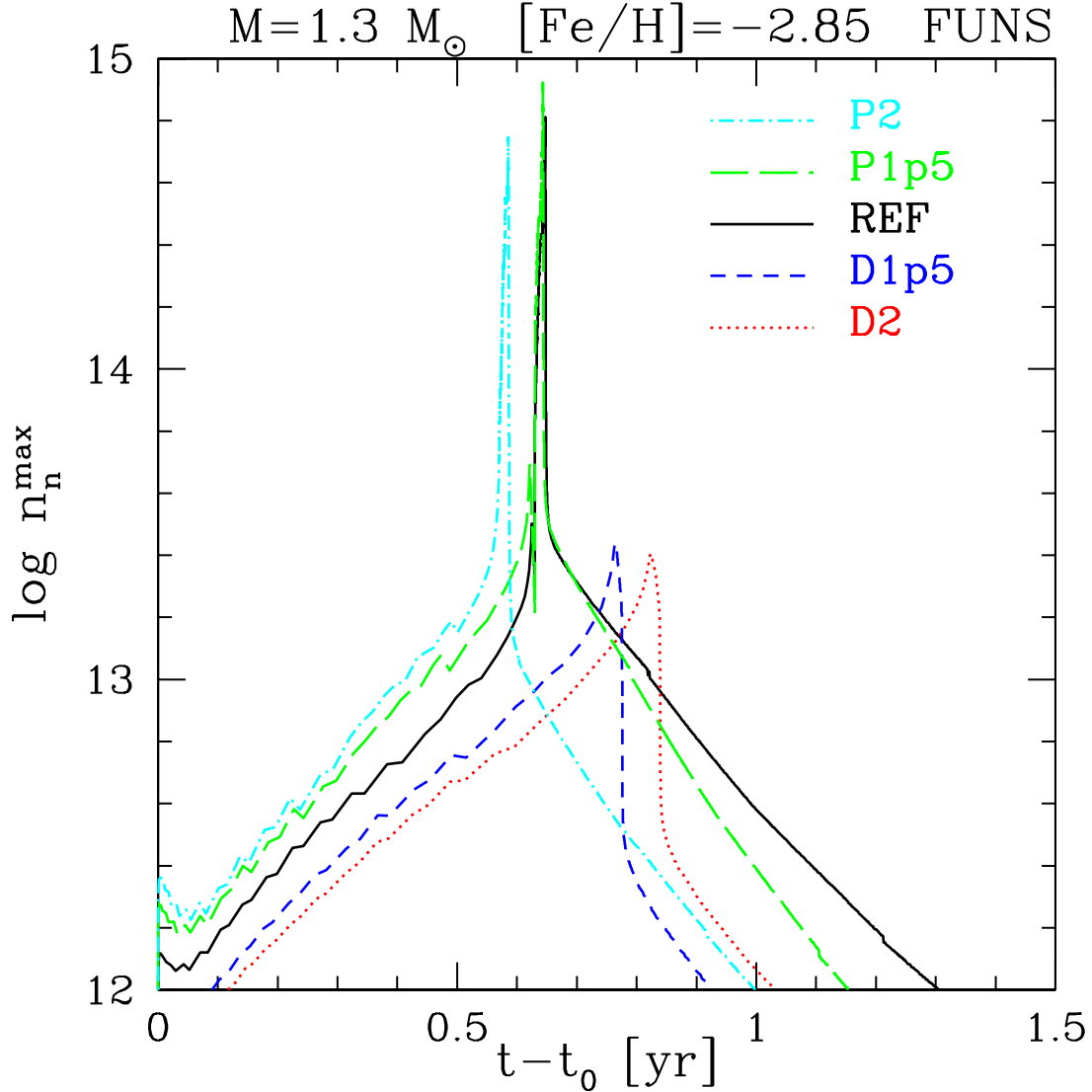}
\caption{Neutron densities attained during the
Proton Ingestion Episode of a star with M= 1.3 M$_\odot$ and
[Fe/H]=-2.85, as a function of different rates for the
$^{13}$C($\alpha$,n)$^{16}$O reaction.} \label{fig_theo_9}
\end{figure*}

\citet{stra14} firstly demonstrated that the $s$-process enriched
ejecta of an early generation of massive AGBs may be responsible
for the s-rich distributions observed in samples of red giant
stars belonging to the globular clusters M4 and M22 (see also
\citealt{shi14}). Those authors found that a satisfactory match to
observations can be obtained, with some noticeable exceptions. In
particular, theoretical models produce too much lead with respect
to observations. \cite{MAS17} highlighted that a variation of the
$^{22}$Ne($\alpha$,n)$^{25}$Mg rate may soften the current
disagreement between theory and observations. Here, we check if
the variation of the $^{13}$C($\alpha$,n)$^{16}$O rate may have
some effects on lead (mainly $^{208}$Pb) production. To that
purpose, we calculate an AGB model with initial mass M= 4M$_\odot$
and [Fe/H]= -2.15. Objects like this are thought to be the typical
polluters for those stellar systems. We find that the model is not
sensitive to any variation of the rate (see Figure
\ref{fig_theo_7}). This result confirms the trend already
highlighted for the M= 3 M$_\odot$ model at higher metallicity
(note that for this model M$_{\rm H}\sim 0.86$ M$_\odot$).
\begin{figure*}[htpb]
\centering
\includegraphics[width=\textwidth]{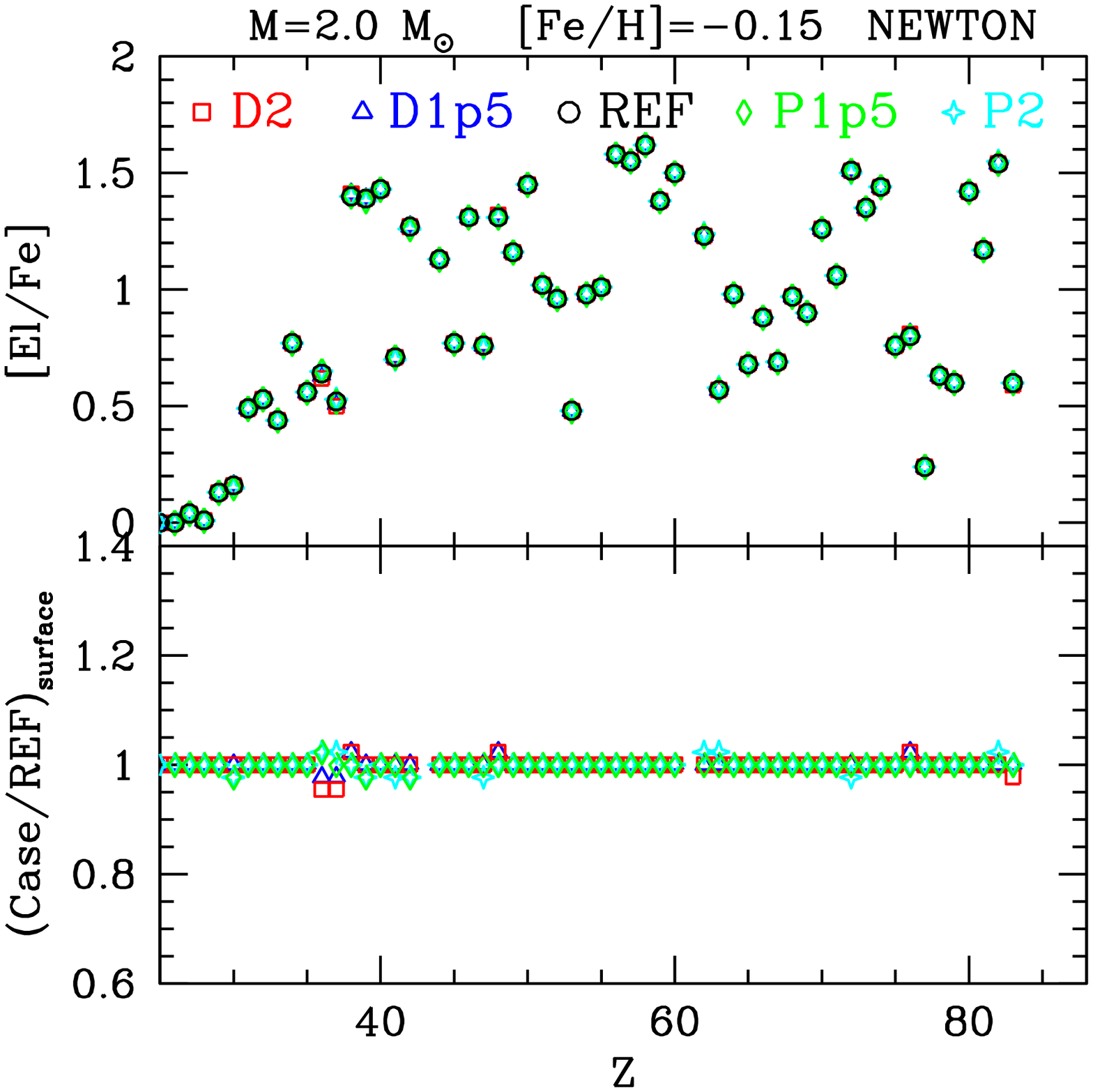}
\caption{As in Figure \ref{fig_theo_1}, but for a
star with initial mass M= 2 M$_\odot$ and [Fe/H]=-0.15.}
\label{fig_theo_10}
\end{figure*}
\begin{figure*}[tpb]
\centering
\includegraphics[width=\textwidth]{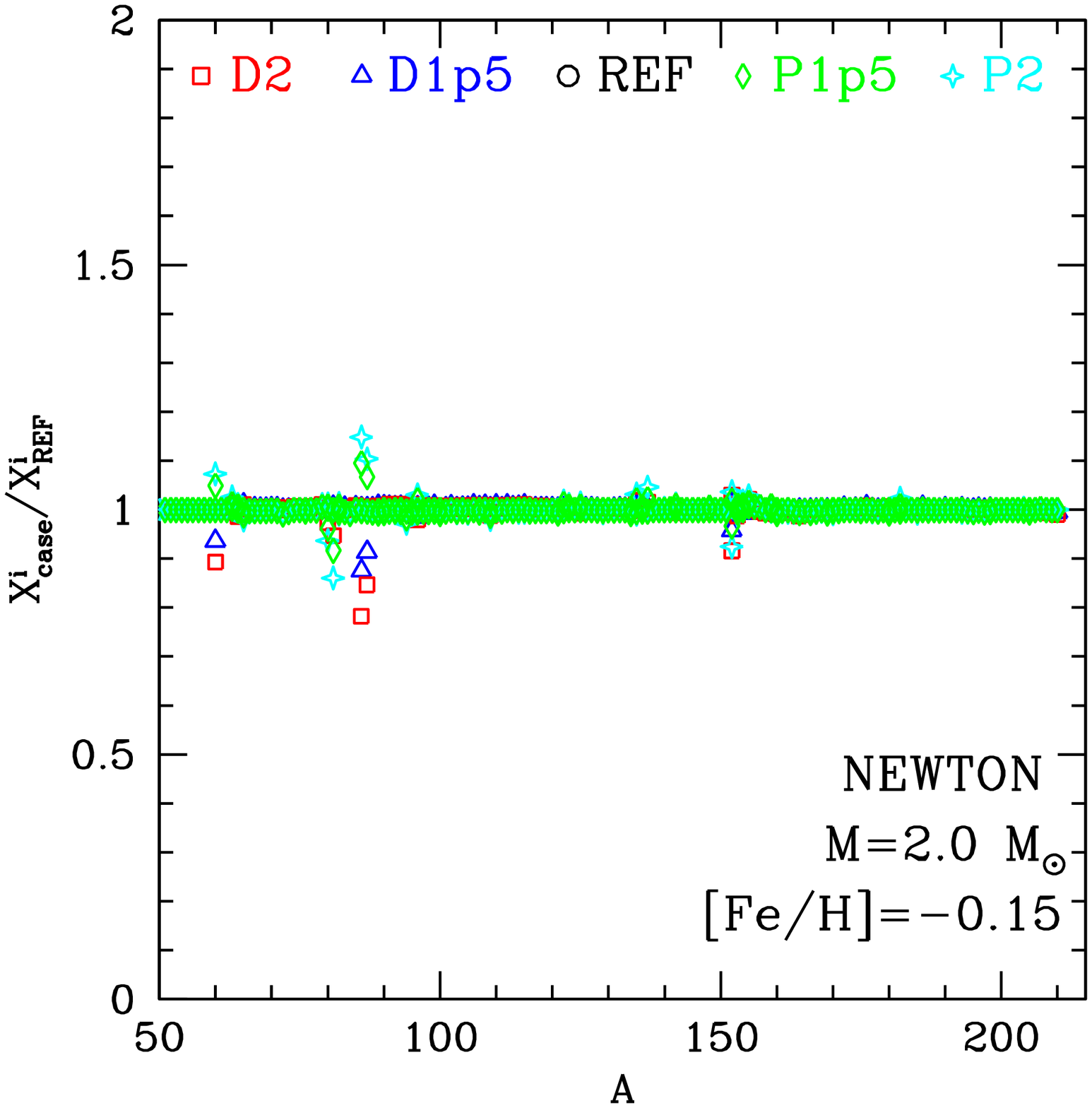}
\caption{As in Figure \ref{fig_theo_10}, but for
the isotopic composition.} \label{fig_theo_11}
\end{figure*}

The situation is definitely different if protons are mixed within
the convective shell generated by a TP. Those type of events may
occur at the first fully developed TP of low-mass low-metallicity
stars or during a very late TP near the tip of the AGB (Sakurai's
objects). As representative of this class of events, we analyze a
low mass (M= 1.3 M$_\odot$) model at very low metallicity (e.g.
[Fe/H]=-2.85). At this Z, the H-shell entropy barrier is weaker
and, as a consequence, some hydrogen may be engulfed in the
growing convective shell triggered by a TP. Due to the large
temperatures, protons are captured by the abundant $^{12}$C while
they are mixed. On the contrary, the corresponding products
($^{13}$C and, eventually, $^{14}$N), reach the base of the
convective shell, where T$\sim$230 MK. Thus, $^{13}$C is exposed
to a temperature well beyond the threshold activation for the
$^{13}$C($\alpha$,n)$^{16}$O reaction ($\sim$ 100 MK). As a
consequence, an efficient $s$-process takes place. After some
months, the energy deposited by the on-flight burning of protons
leads to the splitting of the convective shell. From that moment
on, the upper shell is triggered by an incomplete CNO burning,
while the lower shell is sustained by the 3$\alpha$ and the
$^{13}$C($\alpha$,n)$^{16}$O reactions. \citet{cri09b}
demonstrated that, in order to properly model such a proton
ingestion (named by those authors Proton Ingestion Episode, PIE),
the physical evolution of the model must be coupled to a full
nuclear network, including all chemical species up to lead (see
also \citealt{cri16}). This derives from the fact that, before the
splitting (and the full development of the TP), the
$^{13}$C($\alpha$,n)$^{16}$O reaction provides a substantial
fraction of the local energy budget. This reaction, in fact,
directly releases about 2.2 MeV. Furthermore, an additional energy
contribution comes from the following neutron capture (5 MeV on
average). It is therefore obvious that any model aiming at
following a PIE cannot overlook the proper calculation of the
energetics provided by neutron captures. The number of available
neutron depends on two factors: the mixing efficiency and the
burning efficiency. In this paper, we test the latter by modifying
the rate of the $^{13}$C($\alpha$,n)$^{16}$O reaction. 

However, before discussing our results, some important remarks about mixing have to be highlighted. First, it has to be stressed that the
simulation of a 3-dimensions hydrodynamic event, as a PIE, in a one-dimension
hydrostatic code intrinsically implies the adoption of approximations. A key quantity in stellar evolution is the Damk\"ohler Number $Da=\tau_{mix}/\tau_{burn}$ between the characteristic timescales of convective mixing ($\tau_{mix}$) and nuclear burning ($\tau_{burn}$). In the large majority of steady burning phases of a stellar evolution, $Da << 1$ inside convective zones. During a PIE, instead, $Da \rightarrow 1$, i.e. nuclear burning occurs ‘on-the-fly’ (at least for hydrogen). As a first consequence, the model timestep ($\Delta t$) needs to be reduced to follow the in-flight burning. In our models, we limit the timestep to 50\% of the mixing turnover timescale of the convective shell ($\tau_{mix}$), in order to avoid the unrealistic fully homogenization of that region. All isotopes are mixed within the convective region, apart from protons, which are mixed down to the mass coordinate where $\tau_{burn}=1/3\cdot\Delta t$ (see \citealt{cri09b} for details). At deeper coordinates, $Da >> 1$, i.e. the protons reduction (via nuclear burning) is much more rapid than the supply (via convective mixing). As a consequence, below this point it is hardly unlikely for protons to survive. Other approaches, however, may be implemented, possibly leading to different results.
For instance, \cite{simon} followed a diffusive approach in their calculation of PIEs. Being diffusion a very efficient mixing mechanism, the adoption of a diffusion equation in their code lead to an early splitting of the convective region. This has strong consequences on the following {\it s}-process nucleosynthesis (strongly hampering it).\\
Secondly, we stress that in our code the rate of mixing is calculated basing on the Mixing Length Theory (MLT). This is a very crude approximation,
since the MLT is a local theory and cannot capture the 3-dimensions hydrodynamic nature of a PIE event. In particular, the inhomogeneities in the flow and its effect on the 3-dimension distribution of burning cannot be taken into account. Moreover, MLT cannot treat macroscopic motions, as the Global Non-spherical Oscillations (GOSH) for H-ingestion flashes described by \cite{heidro}. Due to the huge computer time requested, those simulations can be used to constrain 1-dimension hydrostatic calculations, but they cannot substitute them yet. Then, we reckon that explorations as the one presented here still herald precious hints on the physics of PIEs and, thus, we proceed in describing our result.\\
In Figure
\ref{fig_theo_8} we report the heavy element surface distributions
of a 1.3 M$_\odot$ model with [Fe/H]=-2.85. A dichotomy clearly
emerges.
\begin{figure*}[htpb]
\centering
\includegraphics[width=\textwidth]{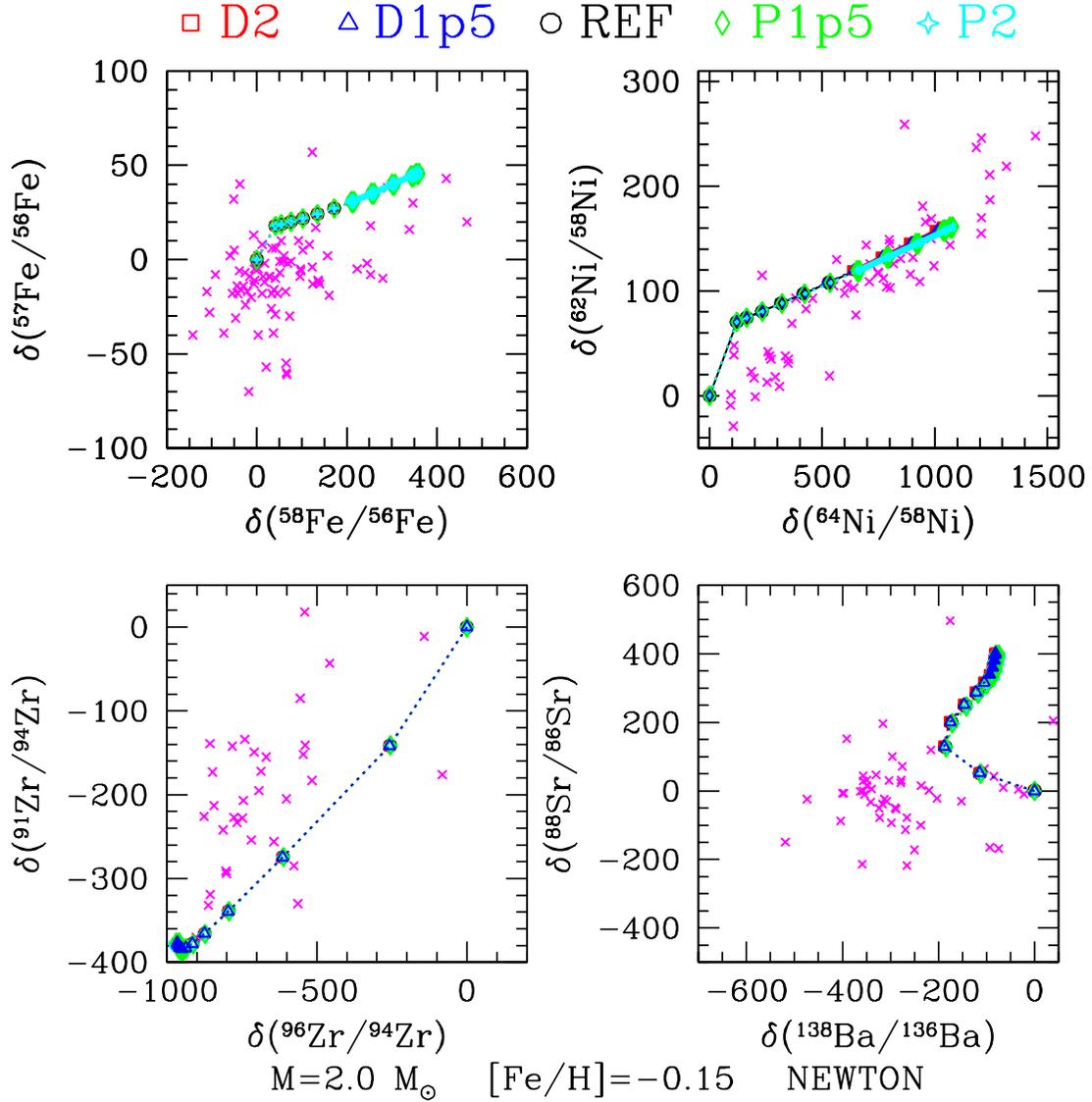}
\caption{As in Figure \ref{fig_theo_10}, but for
the isotopic composition.} \label{fig_theo_12}
\end{figure*}
The 3 cases with an high $^{13}$C($\alpha$,n)$^{16}$O rate (P2,
P1p5 and REF) show similar distributions, even if with important
exceptions (see, e.g., lead and bismuth). On the contrary, other
cases (D1p5 and D2) display a completely different behavior, with
a definitely lower production of heavy elements (more than a
factor 50). Thus, it looks that the variation of the
$^{13}$C($\alpha$,n)$^{16}$O rate induces a sort of threshold
effect in the $s$-process nucleosynthesis of this model. This is
well understood in the framework of the PIE mechanism. Once the
shell has split, the nucleosynthesis of the two shells follow
separate evolutions. The lower one consumes its $^{13}$C
reservoir: for a while in a convective environment and later on,
when convection has switched off, in a radiative way. The upper
shell, instead, continuously ingests protons, producing a large
amount of $^{13}$C, and of $^{14}$N as well (which act as a
poison).
\begin{table}[ht!]
\renewcommand{\thetable}{\arabic{table}}
\centering \caption{$s$-process indexes of the 1.3 M$_\odot$ model
with [Fe/H]= -2.85 for different values of the
$^{13}$C($\alpha$,n)$^{16}$O reaction rate.} \label{lowz}
\begin{tabular}{lcc}
\tablewidth{0pt}
\hline
\hline

Case & [hs/ls] & [Pb/hs] \\
\hline
D2 & -0.33 & 0.63\\
D1p5 & -0.49 & 0.66\\
REF & 0.13& 0.84\\
P1p5 & 0.23& 0.63\\
P2 & 0.21& 0.29\\
\hline
\end{tabular}
\end{table}
As a consequence, the production of heavy elements in the upper
shell almost freezes after the splitting. As soon as the H-burning
switches off, the convective envelope penetrates inward up to the
splitting coordinate, carrying to the surface the material
processed in the upper shell. On the contrary, the $s$-process
rich material of the lower shell is diluted in the following TP
and it is mixed to the surface at the epoch of the following TDU
episode. Later, the model follows a standard AGB evolution. The
two cases with a reduced $^{13}$C($\alpha$,n)$^{16}$O rate ingest
protons as well, but the splitting occurs before enough $^{13}$C
has been mixed to the bottom of the convective shell (and, thus,
an efficient $s$-process nucleosynthesis has developed). In Figure
\ref{fig_theo_9} we report the maximum neutron densities attained
at the base of the convective shell for the different adopted
rates as a function of the time since the beginning of the PIE. It
is evident that the D1p5 and the D2 cases attain lower neutron
densities ($\sim 10^{13}$ cm$^{-3}$ compared to $\sim 10^{15}$
cm$^{-3}$) and, consequently, the following $s$-process
enhancement is lower. The corresponding $s$-process indexes
[hs/ls]\footnote{[hs/ls]=[hs/Fe]-[ls/Fe], where
[ls/Fe]=([Sr/Fe]+[Y/Fe]+[Zr/Fe])/3 and
[hs/Fe]=([Ba/Fe]+[La/Fe]+[Nd/Fe]+[Sm/Fe])/4.} and [Pb/hs] are
reported in Table \ref{lowz} (see discussion in \S
\ref{sec:conclu}).

\subsection{NEWTON models}

The NEWTON code calculates $s$-process nucleosynthesis for low
mass AGB stars considering contributions of both the
$^{13}$C($\alpha$,n)$^{16}$O neutron source and the
$^{22}$Ne($\alpha$,n)$^{25}$Mg one. It uses a network with 404
nuclei (from hydrogen to bismuth). As already highlighted in the
Introduction, the formation of the $^{13}$C pocket is assumed to
be induced by the stellar dynamo. By using the approach proposed
by \citet{nb14}, the proton penetration into the He-rich layers
results to be the consequence of the advection to the envelope of
magnetized material from inner radiative layers: as a consequence,
the $^{13}$C reservoir formed is large (up to $5 \times
10^{-3}$M$_{\odot}$) with an almost flat profile. Such a profile
is quite different from the exponentially-decreasing trend assumed
in several other models (see e.g. the EVEP mixing), while the
$^{14}$N abundance is high just in a very thin layer adjacent to
the envelope. \citet{trippa} firstly showed that the neutron
source  $^{13}$C formed by MHD processes can account for the
abundances of the main s-component nuclei in solar proportions as
well as the abundance distribution of post-AGB objects. In the
same way, \citet{pa18} demonstrated that the magnetic mixing
model, applied to a set of low mass AGB stars (with mass from 1.5
to 3 M$_{\odot}$ and metallicity from 1/3 to 1 Z$_{\odot}$),
provides a satisfactory explanation for s-element isotopic mix
measured in presolar SiC grains.

As done in previous section, we discuss here the effect induced on
NEWTON nucleosynthesis predictions by a change of the
$^{13}$C($\alpha$,n)$^{16}$O reaction rate on a model of 2.0
M$_{\odot}$ and [Fe/H]= -0.15\footnote{Note that the effects of a
variation of the $^{13}$C($\alpha$,n)$^{16}$O reaction in the 1.5
and 3.0 M$_\odot$ models have already been presented by
\cite{TRI17}}. The reaction rates adopted  for the
$^{13}$C($\alpha$,n)$^{16}$O reaction (\citealt{HEI08} divided and
multiplied for  a factor 1.5 and 2) and the notation are the same
of previous section as well as the SiC grain data used for
comparison.

In Figure \ref{fig_theo_10} and in Figure \ref{fig_theo_11} the
heavy element surface composition and the corresponding isotopic
distribution of the 2.0 M$_{\odot}$ AGB star considered are
reported for the five different choices of  the
$^{13}$C($\alpha$,n)$^{16}$O reaction rate. Both the surface
elemental composition and the isotopic one reveal no sensitivity
to the $^{13}$C$+\alpha$ cross section adopted in calculations,
but for a few nuclei, namely $^{60}$Fe, $^{81}$Kr, $^{86}$Kr,
$^{87}$Rb, and $^{96}$Zr, whose abundance variations in any case
range between 5\% and 20\%. The lower sensitivity of NEWTON to the
$^{13}$C($\alpha$,n)$^{16}$O reaction rate with respect to FUNS
models is due to the fact that the $^{13}$C in the pocket is
entirely consumed during the interpulse period. Thus, none is left
to be burnt in the convective shell triggered by the following TP,
for any choice of reaction rate. This is ascribed to the different
stellar structure adopted to compute the post-process calculation,
with respect to the FUNS models described in previous section.
NEWTON models, in fact, use physical inputs computed with the
FRANEC code \citep{stra03}, in which the exponentially decaying
profile of convective velocities was not implemented. Therefore,
the core masses M$_{\rm H}$ at the first TP followed by TDU is
larger for any computed model. For instance, in the 2.0
M$_{\odot}$ studied here M$_{\rm H}\sim 0.61$, to be compared with
the typical values characterizing FUNS models (M$_{\rm H}\sim
0.56$; see previous section). In Figure \ref{fig_theo_12} we
repeat the comparison between isotopic ratios found in pre-solar
SiC grains and the theoretical predictions of the NEWTON code for
the $s$-process nucleosynthesis of the 2.0 M$_{\odot}$ AGB model.
No remarkable changes are yielded by the
$^{13}$C($\alpha$,n)$^{16}$O reaction rate adopted in calculation.
However we highlight the different trends of the curves in this
Figure and the ones of Figure \ref{fig_theo_3} and
\ref{fig_theo_6}, which do not depend on the nuclear physics
input, but on the physical mechanism responsible for the $^{13}$C
pocket formation.

\section{Future perspectives} \label{future}

Theoretical results presented in previous Section, as well as the discussion on the experimental status of the $^{13}{\rm C}(\alpha,$n$){}^{16}{\rm O}$ reaction reported in \S \ref{sec:expe}, clearly call for further direct
measurements of its S-factor within the Gamow window corresponding to a radiative $^{13}$C burning. Owing to the vanishingly small cross section, it
is then very likely that only underground measurements can help to
sort out the discrepancies affecting direct data on the $^{13}{\rm
C}(\alpha,$n$){}^{16}{\rm O}$ reaction, provided that the neutron
detection efficiency of the setup is accurately known. The
application of underground facilities to its investigation is
detailed in Sect.\ref{luna}.

Alternative approaches would also help to clarify the
contradictory results brought by direct and indirect methods. In
particular, the use of the detailed balance theorem
(Sect.\ref{ntof}) might be of great help as an independent
indirect technique, since the $^{13}{\rm C}(\alpha,$n$){}^{16}{\rm
O}$ cross section could be obtained with no need of neutron
detectors, reducing possible sources of systematic errors linked
to n-detection.

\subsection{The LUNA experiment} \label{luna}

In order to tackle the very challenging direct measurement of the
low-energy cross section the signal-to-background ratio needs to
be as high as possible. The main limitation of the previous
experiments that went the lowest in energy \citep{DRO93, HEI08}
was the natural background radiation, which at low enough energies
(and with that, cross sections) becomes too dominant to allow
further continuation towards the Gamow peak for a radiative
$^{13}$C burning.

Since the detection efficiency of the setups and the beam currents
impinging the targets were already rather high, only small signal
rate gains of maybe up to an order of magnitude can be achieved by
improving these aspects of the measurement. A further improvement
on the state of the art of direct measurements seems to be only
possible through a drastic suppression of the environmental backgrounds.\\
In deep underground laboratories like the Gran Sasso National
Laboratories (LNGS) both the cosmic-ray induced $\gamma$-ray and
neutron backgrounds are drastically reduced with respect to the
surface: $\gamma$s by about six and (thermal) neutrons by up to
three orders of magnitude \citep{BES16a, BES16b}. The Laboratory
for Underground Nuclear Astrophysics (LUNA) at the LNGS has for
over 25 years exploited the low-background conditions underground
to measure astrophysical relevant nuclear reaction cross sections
close to or directly inside the relevant stellar burning energies
(see \citealt{BES16b} and references therein).
The LUNA 400 accelerator \citep{FOR03} can provide
50-400 keV proton and alpha beams with currents up to around 400
$\mu$A, covering the energy range of interest for this reaction.
To successfully measure the very low cross sections in the
astrophysical energy range LUNA will use a high-efficiency detector
made of $^3$He counters embedded in a moderating polyethylene
matrix (a proven design used in many experiments in the past, see e.g.
\citet{FAL13} and references therein). At the low expected count
rates (a few neutrons per day at the lowest energies) the internal
alpha activity of the counters themselves becomes a relevant 
background\citep{HAS98}. It will be suppressed using pulse-shape
discrimination methods \citep{LAN13}, further improving the
signal-to-background ratio.

The measurement of $^{13}$C$(\alpha, n)^{16}$O is also in the list
of reactions to be measured with a second, high-energy accelerator
to be installed at LNGS, LUNA MV. The energy range covered by this
new device is from 200 kV to 3.5 MV, providing an overlap region
with LUNA 400 keV (the accelerator currently used). LUNA MV will
connect the low-energy data to the higher-energy region, allowing
an additional cross-check of the systematic uncertainties. The
extended data set will also provide very valuable for a global
analysis using, for example, an R Matrix approach.

In summary, the LUNA experiment aims at a direct measurement of
the cross section of the $^{13}{\rm C}(\alpha,$n$){}^{16}{\rm O}$
reaction at the Gamow-energy for radiative $^{13}$C burning. Based
on preliminary studies of the intrinsic activity of the detector
and the laboratory background it appears realistic to map out the
reaction down to $\approx$ 250 keV(c.m.). The lowest-energy data
points will - after a few weeks of data taking - reach precisions
high enough to much better constrain the cross section in the
Gamow energy window than it currently is. The reduction of the
uncertainty of the reaction rate will help, in combination with
other direct, from LUNA MV and other sources, and indirect data,
to resolve the open astrophysical questions outlined in the
introduction of this paper.

\subsection{The n\_TOF experiment} \label{ntof}
As already discussed above, when the direct approach to the study
of ($\alpha$,n) reactions is particularly difficult, indirect or
inverse reactions are considered as a valid method for
constraining the reaction cross sections of astrophysical
interest. For instance at the n\_TOF facility at
CERN~\citep{GUE13} an important contribution to the study of the
$^{22}$Ne($\alpha$, n)$^{25}$Mg neutron source was recently
provided, on the basis of neutron spectroscopy of $^{26}$Mg
states~\citep{MAS17}.

The challenging $^{13}{\rm C}(\alpha, $n$){}^{16}{\rm O}$
measurement can benefit from experimental information from the
time-reversed reaction $^{16}{\rm O}(n, \alpha){}^{13}{\rm C}$. In
particular, by using the detailed balance (i.e. time-reversal
invariance theorem), the reaction cross section of the $^{13}{\rm
C}(\alpha, $n$){}^{16}{\rm O}$ is deduced from the measurement in
the reverse direction.
As shown in Fig.~\ref{fig:ReactionScheme}, the nuclear reaction is
a two-step process: first an excited state of the $^{17}$O
compound nucleus is populated and, after, it decays into an exit
channel. Since any excited state is characterized by its spin and
parity, another interesting relation between direct and inverse
reaction arises: the value of the channel spin of the entrance
($^{16}$O+n) and the exit channel ($^{13}$C+$\alpha$) is the same,
$\hbar$/2, but the parity is opposite. As a consequence, if the
formation of a resonant state is possible via $s-$wave neutrons
($\ell =0$), it cannot be formed via $s-$wave $\alpha$ particles
and it can occur most probably via $p-$wave.  In summary, beside the Coulomb barrier, which strongly suppresses
the cross section of the direct reaction at low energy and does
not affect the inverse reaction, also the orbital angular momentum
can play a role in the suppression or enhancement of the
resonance-cross section.
The $\alpha$-unbound states of
$^{17}$O can be studied by impinging a neutron beam on a $^{16}$O
target. Because of the difference of about 2.22 MeV in the
$Q$-value of direct and inverse reaction, the threshold
neutron-energy of the $^{16}{\rm O}(n, \alpha){}^{13}{\rm C}$
reaction is about 2.35 MeV. This kind of measurement will be performed at the neutron
time-of-flight facility of CERN (n\_TOF). In particular, thanks to
its excellent energy resolution and high neutron flux, the n\_TOF
facility offers the opportunity to perform such a measurement at a
sufficiently large number of energies and to resolve fine
structures in the cross section. The n\_TOF EAR1 experimental
area is placed at a distance of about 185-m from the spallation
target, and it is best suited for high-precision measurements
thanks to its excellent energy resolution ($\Delta E/E=5.3\times
10^{-3}$ at $E_n=1$ MeV) which allows to perform a precise
resonance shape analysis.

The detection setup consists in a double Frisch-grid ionization
chamber with common cathode. All the material in-beam has been
kept to the minimum so to reduce the effect due to the in-beam
$\gamma$-ray burst generated by the spallation reactions. The gas
mixture is composed of Kr(95\%) + CO$_2$(5\%), where the latter
acts as the oxygen sample itself. The use of an ionization chamber
with the gas acting also as target allows one to detect $\alpha$
particles with energy as low as a few hundred keV, since the full
particle energy is released inside the active volume of the
detector. As a consequence, it may well be possible with this
technique to observe and characterize the first few levels above
the $\alpha$-threshold. For instance, the time-reversal
measurement planned at the n\_TOF facility may provide additional
information on the $5/2^+$ (at $E_x=7.164$ MeV, corresponding to
$E_n=3.21$ MeV) and $7/2^-$ (at $E_x=7.318$ MeV, corresponding to
$E_n=3.44$ MeV) state, corresponding to $\alpha$-particle energies
of 500 and 700 keV.

\section{Discussion and Conclusions} \label{sec:conclu}

The sensitivity study performed in this work herald interesting
results, which further strengthen the need of more detailed
nuclear data concerning the $^{13}$C($\alpha$,n)$^{16}$O cross
section. In order to evaluate the effect on $s$-process
nucleosynthesis induced by the use of any of the
$^{13}$C($\alpha$,n)$^{16}$O rates currently available in the
literature, our study focused on a theoretical range well beyond
the most recent uncertainty estimates. First, we carefully scrutinized the pool of experimental data regarding the $^{13}$C($\alpha$,n)$^{16}$O cross section to find out the uncertainty affecting the corresponding reaction rate. Owing to the large scatter in direct measurements, discrepancies as large as a factor of 2 are apparent in the absolute value of the cross section and then in the calculated reaction rate (see \S \ref{sec:expe}), while an average error of ~25\% have been calculated taking into account all available data sets (see \citealt{TRI17}). Therefore, we varied by a factor 1.5 and 2 (upward and downward) the rate proposed by \cite{HEI08}, assumed as a reference case, to include potential systematic errors. Most interesting
results are:
\begin{itemize}
\item{a variation of the $^{13}$C($\alpha$,n)$^{16}$O rate does
not appreciably affect $s$-process distributions for masses above
3 M$_\odot$ at any metallicity. The results obtained with the
NETWON post-process code and the FUNS evolutionary code are
consistent among them. Apart from a few isotopes, the differences
deriving from a variation of the $^{13}$C($\alpha$,n)$^{16}$O
reaction are always below 5\%. This basically confirm the previous
finding by \cite{GUO12} and \cite{TRI17};} \item{the situation is
completely different if the standard paradigm of the $s$-process
(i.e. that all $^{13}$C within the pockets burns radiatively) is
violated. This occurs in FUNS models of low mass (M$<$3 M$_\odot$)
at solar-like metallicities (\citealt{cri09a}, see also
\citealt{ka10}). In such a case, a change of the
$^{13}$C($\alpha$,n)$^{16}$O reaction rate leads to non-negligible
variations in both the elemental and isotopic composition of the
model. On average, elements surface distribution differ by about
10\% with respect to the reference model, the heavier species
showing the largest variations. A peak of 30\% is obtained for
rubidium: this is a consequences of the large overproduction
(underproduction) of $^{87}$Rb for slower (faster)  reaction
rates. Such a trend is also found for other neutron-rich isotopes
as, for instance, $^{60}$Fe (a factor 20). Unfortunately, typical
uncertainties affecting the spectra of $s$-process rich stars are
beyond the expected theoretical variations. Our analysis can
safely discard extremely low values for the
$^{13}$C($\alpha$,n)$^{16}$O rate only. More precise constraints
can be extracted from pre-solar SiC grains data. A change of the
$^{13}$C($\alpha$,n)$^{16}$O reaction rate within the explored
range produces substantial differences in the isotopic surface
ratios, in particular for the zirconium isotopes. However, the
intrinsic spread characterizing SiC grains does not allow to draw
any firm conclusion. Larger $^{13}$C($\alpha$,n)$^{16}$O rates
also produce larger surface $^{86}$Kr/$^{82}$Kr isotopic ratios.
Even if this result is going in the right direction, the extreme
ratios ($\sim$3) found by \cite{verco} cannot be matched.
Therefore, we confirm the results by \cite{GUO12};} \item{the by
far more interesting comes from low-mass low-metallicity FUNS
models (M= 1.3 M$_\odot$ with [Fe/H]= -2.85). In this case, during
the first fully developed Thermal Pulse, some protons are engulfed
in the underlying convective shell and burn on-fly (Proton
Ingestion Episode). During this peculiar phase, characterized  by
an extraordinary rich nucleosynthesis, the energy budget results
from the balance between H-burning and He-burning. The latter
receives an important energetic contribution from the
$^{13}$C($\alpha$,n)$^{16}$O reaction. Depending on the adopted
rate, completely different results are attained. In particular,
the surface abundances of the heavier elements may decrease by
more than a factor 50, when slow rates for the
$^{13}$C($\alpha$,n)$^{16}$O reaction are used. Our results are
substantially different from those obtained by \cite{GUO12}. The
reason lies in the different approaches to the modeling of proton
ingestions. In \cite{GUO12} work, a post-process technique has
been applied to follow the on-going $s$-process nucleosynthesis.
In such a case, any energetic feedback from the
$^{13}$C($\alpha$,n)$^{16}$O reaction and, most important, from
the following neutron capture, is lost. It has been demonstrated
in the past that the physical evolution of the model strongly
depends on that. The difference in the results follows
consequently. Unfortunately, also in this case observations cannot
help in constraining the $^{13}$C($\alpha$,n)$^{16}$O rate. In
fact, halo stars showing $s$-process enhanced distributions owe
their surface compositions to pollution events from an already
extinct AGB. From the pollution episode up to now, other mixing
processes may have modified the observed distribution (such as
gravitational settling, see e.g. \citealt{sta07}). Thus, we cannot
take advantage of absolute abundances to derive firm conclusions.
In principle, this problem can be circumvented by looking at the
relative enhancement of the three $s$-process peaks. Observations
indicate that, on average, [hs/ls]$>$0.3 (see, e.g. Fig. 5 in
\citealt{cri16}). Those values are consistent with the numbers
obtained with the highest $^{13}$C($\alpha$,n)$^{16}$O reaction
rates (REF, P1p5 and P2; see Table \ref{lowz}). On the other hand,
there are some isolated stars showing very low [hs/ls] values
($\sim -0.5$), which would be consistent with low rates (cases
D1p5 and D2; see Table \ref{lowz}). Thus, if we base our reasoning
on a mere probabilistic criterion, we should exclude the lowest
rates. However, it has to be taken into account that the
calculation of a proton ingestion episode in a one-dimension
hydrostatic code involves many approximations, which are hard to
be verified (see, e.g., \citealt{heidro}). Moreover, the
occurrence itself of this peculiar type of mixing in low-mass
low-metallicity stars is still matter of debate, since it strongly
depends on the initial composition (and, in particular, on the
initial enrichment of $\alpha$ elements). In conclusion,
regardless of the robustness of the obtained results, we cannot
get any hints on the efficiency of the
$^{13}$C($\alpha$,n)$^{16}$O reaction in those stars.}
\end{itemize}
Current observational and laboratory uncertainties cause our
attempt to constrain the $^{13}$C($\alpha$,n)$^{16}$O rate by
means of stellar models to be almost fruitless. In the future,
however, very high-resolution spectrographs mounted on next
generation telescopes (e.g. HIRES on the E-ELT telescope;
\citealt{marconi}) as well as advanced resonant ionization mass
spectrometers (e.g. CHARISMA; \citealt{savina}) will provide
extremely precise data (reducing the related errors and
disentangling an eventual contamination from  solar system
material, respectively). In the meanwhile, any improvement in our
knowledge of the $^{13}$C($\alpha$,n)$^{16}$O rate will be useful.
The ongoing experimental effort, illustrated in this work, is
expected to produce in the next years a new set of direct and
indirect data on the $^{13}$C($\alpha$,n)$^{16}$O reaction. The
determination of its reaction rate will benefit from the crucial
information from the direct measurement, and will take advantage
of the additional constraints from indirect measurements (e.g. the
strong sensitivity of THM to the $^{17}$O level near the
$\alpha-$threshold). In addition, the combination of the direct
$^{13}$C($\alpha$,n)$^{16}$O data from LUNA and the inverse
$^{16}$O($n,\alpha$)$^{13}$C data from n\_TOF can be used to
reduce any possible source of systematic uncertainties. In
summary, an accurate characterization of the resonant structures
in the $^{13}$C($\alpha$,n)$^{16}$O reaction cross section in the
energy region of interest can be attained. It will eventually
constraint the reaction rate to a conclusive level, well below the
limits considered in the present study.

\acknowledgments

SP acknowledges the support of Fondazione Cassa di Risparmio di
Perugia.


\begin{thebibliography}{}
\bibitem [Abia et al.(2001)]{ab01}Abia, C., et al., 2001, \apj, 559, 1117
\bibitem[Astropy Collaboration et al.(2013)]{2013A&A...558A..33A} Astropy Collaboration, Robitaille, T.~P., Tollerud, E.~J., et al.\ 2013, \aap, 558, A33
\bibitem[Avila et al.(2015)]{AVI15}Avila, M.L., et al., 2015, \prc, 91, 048801
\bibitem[Bair \& Haas(1973)]{BAI73}Bair, J.K., and Haas, F.X., 1973, \prc, 7, 1356
\bibitem[Battino et al.(2016)]{ba16}Battino, U., et al., 2016, \apj, 827, 30
\bibitem[Best et al.(2016a)]{BES16a}Best, A. et al., 2016, Nucl. Inst. Meth. A, 812, 1
\bibitem[Best et al.(2016b)]{BES16b}Best, A. et al., 2016, The European Physical Journal A, 52, 72
\bibitem[Bisterzo et al.(2015)]{bi15}Bisterzo, S., et al., 2015, \mnras, 449, 506
\bibitem[Bracci et al.(1990)]{BRA90} Bracci, L., Fiorentini, G., Melezhik, V. S., Mezzorani, G., \& Quarati, P. 1990, Nucl. Phys. A, 513, 316
\bibitem[Busso et al.(1988)]{bu88}Busso, M., et al., 1988, \apj, 326, 196
\bibitem[Busso et al.(1995)]{bu95}Busso, M., et al., 1995, \apj, 446, 775
\bibitem[Busso et al.(1999)]{bgw}Busso, M., et al., 1999, \araa, 37, 239
\bibitem[Campbell \& Lattanzio (2008)]{simon}Campbell, S.W. \& Lattanzio, J.C., 2008, \aap, 490, 769 
\bibitem[Caughlan \& Fowler(1988)]{cf88}Caughlan, G.R. \& Fowler, W.A., 1988, At. Data Nucl.Tables, 40, 283
\bibitem[Chieffi \& Straniero(1989)]{cs89}Chieffi, S. \& Straniero, O., 1989, \apj, 71, 47
\bibitem[Cristallo et al.(2006)]{cri06}Cristallo, S., et al., 2006, \memsai, 77,
774
\bibitem[Cristallo et al.(2009a)]{cri09a}Cristallo, S., et al., 2009, \apj, 696, 797
\bibitem[Cristallo et al.(2009b)]{cri09b}Cristallo, S., et al., 2009, \pasa, 26, 139
\bibitem[Cristallo et al.(2011)]{cri11}Cristallo, S., et al., 2011, \apjs, 197, 2
\bibitem[Cristallo et al.(2015)]{cri15}Cristallo, S., et al., 2015, \apjs, 219, 21
\bibitem[Cristallo et al.(2016)]{cri16}Cristallo, S., et al., 2016, \apj, 833, 181
\bibitem[Csedreki et al.(2017)]{CSE17}Csedreki, L., et al., 2017, EPJ Web of Conferences, 165, 01017
\bibitem[Davids(1968)]{DAV68} Davids, C. 1968, Nucl. Phys. A, 110, 619
\bibitem [Denissenkov \& Tout(2003)]{dt03} Denissenkov, P.A., \& Tout, C.A., 2003, \mnras,  340, 722
\bibitem[Descouvemont(1987)]{DES87}Descouvemont, P. 1987, \prc, 36, 2206
\bibitem[Drotleff et al.(1993)]{DRO93}Drotleff, H.W., et al., 1993, \apj, 414, 735
\bibitem[Dufour \& Descouvemont(2005)]{DUF05}Dufour, M., \& Descouvemont, P. 2005, \prc, 72, 015801
\bibitem[ENDFS(2017)]{ENDF}International Network of Nuclear Structure and Decay Data Evaluators (http://www-nds.iaea.org/nsdd/)
\bibitem[Falahat et al.(2013)]{FAL13}Falahat, S., et al., 2013, Nucl. Instr. Meth. A, 700, 53
\bibitem[Formicola et al.(2003)]{FOR03}Formicola, A. et al., 2003, Nucl. Instr. Meth. A, 507, 609
\bibitem[Faestermann et al.(2015)]{FAE15}Faestermann, T., et al., 2015, \prc, 92 052802
\bibitem [Freytag et al.(1996)]{fr96} Freytag, B. et al., 1996, {\it A\&A}, 313, 497
\bibitem [Gallino et al.(1988)]{ga88} Gallino, R., et al., 1988, \apjl, 334, 45
\bibitem [Gallino et al.(1998)]{ga98} Gallino, R., et al., 1998, \apj, 497, 388
\bibitem [Goriely \& Mowlavi(2000)]{goriely} Goriely, S. \& Mowlavi, N. ,2000, \aap, 362, 599
\bibitem[Guandalini \& Cristallo(2013)]{guanda} Guandalini, R., \& Cristallo, S., 2013, \aap, 555, 120
\bibitem[Guerrero et al.(2013)]{GUE13}Guerrero, C., et. al., 2013,  Eur. Phys. J. A , 49, 27
\bibitem[Guo et al.(2012)]{GUO12}Guo, B., et al., 2012, \apj, 756, 193
\bibitem[Harissopulos et al.(2005)]{HAR05} Harissopulos, S., Becker, H. W., Hammer, J. W., et al. 2005, PhRvC, 72, 062801
\bibitem[Hashemi-Nezhad et al.(1998)]{HAS98} Hashemi-Nezhad, S. R., Peak, L. S. 1998, Nucl. Instr. Meth. A 416, 100
\bibitem[Heil et al.(2008)]{HEI08}Heil, M., et al., 2008, \prc, 78, 025803
\bibitem [Herwig et al.(1997)]{he97} Herwig, F., et al., 1997, {\it A\&A Lett.}, 324, 81
\bibitem [Herwig et al.(2000)]{he00} Herwig, F., 2000, {\it A\&A}, 360, 952
\bibitem [Herwig et al.(2011)]{herwig} Herwig, F., et al., 2011, {\it ApJ}, 727, 89
\bibitem [Herwig et al.(2014)]{heidro} Herwig, F., et al., 2014, \apjl, 792, 3
\bibitem [Hollowell \& Iben(1988)]{hi88} Hollowell, D., \& Iben, I.Jr., 1988, \apjl, 333,
25
\bibitem [Iben(1975)]{ib75} Iben, I.Jr., 1975, \apj, 196, 525
\bibitem [Iben(1981)]{ib81} Iben, I.Jr., 1981, \apj, 243, 987
\bibitem [Iben \& Renzini(1982)]{ir82} Iben, I.Jr., \& Renzini, A.: 1983, \apjl, 263, 23
\bibitem[Johnson et al.(2006)]{JOH06} Johnson, E. D., Rogachev, G. V., Mukhamedzhanov, A. M., et al. 2006, \prl,
97, 192701
\bibitem[Karakas et al.(2010)]{ka10} Karakas, A., Campbell, S.W., \& Stancliffe, R.J. 2010, \apj, 713, 374
\bibitem[Keeley et al.(2003)]{KEE03} Keeley, N., Kemper, K. W., \& Khoa, D. T. 2003, Nucl. Phys. A, 726, 159
\bibitem[Kellogg et al.(1989)]{KEL89} Kellogg, S., Vogelaar, R., \& Kavanagh, R. 1989, BAPS, 34, 1192
\bibitem[Kubono et al.(2003)]{KUB03}Kubono, S., Abe, K., Kato, S., et al. 2003, \prl, 90, 062501
\bibitem[La Cognata et al.(2010)]{LAC10} La Cognata, M., Spitaleri, C., and Mukhamedzhanov, A.M., 2010, \apj, 723, 1512
\bibitem[La Cognata et al.(2012)]{LAC12} La Cognata, M., et al., 2012, \prl, 109, 232701
\bibitem[La Cognata et al.(2013)]{LAC13} La Cognata, M., et al., 2013, \apj, 777, 143
\bibitem [Lambert et al.(1995)]{la95} Lambert, D.L., et al., 1995, \apj, 450, 302
\bibitem [Langer et al.(1999)]{la99} Langer, N., et al., 1999, {\it A\&A Lett.}, 346, 37
\bibitem[Langford et al.(2013)]{LAN13} Langford, T. J. et al., 2013, Nucl. Instr. Meth. A, 717, 51
\bibitem[Liu et al.(2014a)]{liu14a}Liu, N., et al., 2014, \apj, 786, 66
\bibitem[Liu et al.(2014b)]{liu14b}Liu, N., et al., 2014, \apj, 788, 163
\bibitem[Liu et al.(2015)]{liu15}Liu, N., et al., 2015, \apj, 803, 12
\bibitem[Lugaro et al.(2003)]{lugaro}Lugaro, M., et al., 2003, \apj, 593, 486
\bibitem[Malaney(1987)]{ma87} Malaney, R.A., 1987, \apj, 321, 832
\bibitem[Malaney \& Lambert(1988)]{ml88} Malaney, R.A., \& lambert, D.L., 1988, \mnras, 235, 695
\bibitem[Marconi et al.(2016)]{marconi} Marconi, A., et. al., 2016, SPIE, 9908, 23
\bibitem[Massimi et al.(2017)]{MAS17} Massimi, C., et. al., 2017, Phys. Lett. B, 768, 1
\bibitem[Mezhevych et al.(2017)]{MEZ17}Mezhevych, S.Yu., et al., 2017, \prc, 95, 034607
\bibitem[Mostefaoui et al.(2005)]{moste} Mostefaoui, S., et al., 2005, \apj, 625, 271
\bibitem[Mukhamedzhanov et al.(2017)]{MUK17}Mukhamedzhanov,  A. M., et al., 2017, \prc 96, 024623
\bibitem[Nucci \& Busso(2014)]{nb14} Nucci M. C. \& Busso M., 2014, \apj 787, 141.
\bibitem[Palmerini et al.(2018)]{pa18}Palmerini, S., et al., 2018, \gca, 221, 21
\bibitem[Parker(1960)]{parker} Parker, E. N.,1960, ApJ, 132, 821
\bibitem[Pellegriti et al.(2008)]{PEL08} Pellegriti, M.G., et al., 2008 \prc, 77, 042801
\bibitem[Piersanti et al.(2013)]{pi13} Piersanti, L., et al., 2013 \apj, 774, 98
\bibitem[Raut et al.(2013)]{raut13}Raut, R., et al., 2013, \prl, 111, 112501
\bibitem [Savina et al.(2003)]{savina}Savina, M.R., et al., 2003, \gca, 67, 3215
\bibitem[Shingles et al.(2014)]{shi14}Shingles, L.J., et al., 2014, \apj, 795, 34
\bibitem[Schuessler(1977)]{schu}Schuessler M.,1977, \aap, 56, 439.
\bibitem [Smith \& Lambert(1986)]{sl86} Smith, V.V., Lambert, D.L., 1986, \apj, 311, 843
\bibitem[Spruit(1999)]{spruit} Spruit H. C.,1999, \aap, 349, 189.
\bibitem[Stancliffe et al.(2007)]{sta07}Stancliffe, R., et al., 2007, \aap, 464, 57
\bibitem[Straniero et al.(1995)]{stra95}Straniero, O., et al., 1995, \apjl, 440, 85
\bibitem [Straniero et al.(1997)]{stra97}Straniero, O., et al., 1997, \apj, 478, 332
\bibitem[Straniero et al.(2003)]{stra03}Straniero, O., et al., 2003, \pasa, 20, 389
\bibitem[Straniero et al.(2006)]{stra06}Straniero, O., et al., 2006, \nphysa, 777, 311
\bibitem[Straniero et al.(2014)]{stra14}Straniero, O., et al., 2014, \apj, 785, 77
\bibitem[Shubhchintak et al.(2016)]{subi16}Shubhchintak, et al., 2016, J. Phys. G, 43, 125203
\bibitem[Sugimoto \& Nomoto(1975)]{suno75}Sugimoto, D., \& Nomoto, K., \pasj, 27, 197
\bibitem[Tang \& Dauphas(2012)]{td}Tang, H., \& Dauphas, N., {\it Earth Plan. Sci. Lett.}, 359-360, 248
\bibitem[Tilley et al.(1993)]{TIL93}Tilley, D.R., Weller, H.R., and Cheves, C.M., 1993, Nucl. Phys. A, 564, 1
\bibitem[Trappitsch et al.(2018)]{reto}Trappitsch, R., et al., 2018, \gca, 221, 87
\bibitem[Tribble et al.(2014)]{TRI14} Tribble, R.E., et al., 2014,Reports on Progress in Physics, 77, 106901
\bibitem[Trippella et. al(2016)]{trippa} Trippella, O., et al., 2016, \apj, 818, 125
\bibitem[Trippella \& La Cognata(2017)]{TRI17} Trippella, O., and La Cognata, M., 2017, \apj, 837, 41
\bibitem [Truran \& Iben(1977)]{ti77} Truran, J.W., \& Iben, I.Jr., 1977, \apj, 216, 797
\bibitem[Xu et al.(2013)]{XU13} Xu, Y., Takahashi, K., Goriely, S., et al. 2013, Nucl. Phys. A, 918, 61
\bibitem[Verchovsky et al.(2004)]{verco}Verchovsky, A.B., Wright, I.P., \& Pillinger, C.T. 2004, \apj, 607, 611
\bibitem[Werner \& Herwig(2007)]{wehe}Werner, K., \& Herwig, F., 2007, {\it PASP}, 118, 183

\end{thebibliography}
\end{document}